\documentclass[3p,times]{elsarticle}
\usepackage{ecrc}
\volume{00}
\firstpage{1}
\journalname{Digital Signal Processing}

\runauth{}

\jid{procs}

\jnltitlelogo{Digital Signal Processing}
\usepackage[figuresright]{rotating}

\pdfoutput=1
\usepackage{amsmath,amsthm}
\usepackage{graphics} 
\usepackage{epsfig,epstopdf} 
\usepackage[export]{adjustbox}
\usepackage[dvipsnames]{xcolor}
\usepackage{amsmath,bigints,bbm} 
\usepackage{amssymb}  
\usepackage[T1]{fontenc}
\usepackage[utf8]{inputenc}
\usepackage{cleveref}
\usepackage{caption}
\usepackage{subcaption}
\usepackage{mathtools}
\usepackage{tikz}

\crefrangelabelformat{equation}{(#3#1#4--#5#2#6)}
\crefname{equation}{Eq.}{Eqs.}
\Crefname{equation}{Equation}{Equations}

\newtheoremstyle{problemstyle}  
{3pt}                                               
{3pt}                                               
{\normalfont}                               
{}                                                  
{\bfseries\itshape}                 
{\normalfont\bfseries:}         
{.5em}                                          
{}                                                  
\theoremstyle{problemstyle}

\usepackage{mathtools}

\DeclarePairedDelimiter\floor{\lfloor}{\rfloor}
\usepackage{algorithm}
\usepackage{algorithmic}
\begin{document}

\begin{frontmatter}
	
	
	
	\dochead{}
	
	\title{Sensor Placement and Resource Allocation for Energy Harvesting IoT Networks}
	
	
	\author{Osama M. Bushnaq, 
		Anas Chaaban, 
		Sundeep Prabhakar Chepuri, 
		Geert Leus, 
		Tareq Y. Al-Naffouri
	}

	\address{
		{O. M. Bushnaq and T. Y. Al-Naffouri are with the Electrical Engineering department, King Abdullah University of Science and Technology, Thuwal, Saudi Arabia, (e-mail: osama.bushnaq@kaust.edu.sa; tareq.alnaffouri@kaust.edu.sa). }
		{S. P. Chepuri is with the Department of Electrical Communication Engineering (ECE), Indian Institute of Science, Bangalore - 560 012. India (e-mail: spchepuri@iisc.ac.in). }
		{A. Chaaban is with the School of Engineering, The University of British
			Columbia, Kelowna, BC V1V 1V7, Canada (e-mail: anas.chaaban@ubc.ca). }
		{G. Leus is with the Faculty of Electrical, Mathematics and Computer Science, Delft University of Technology, Delft 2628CD, The Netherlands (e-mail:g.j.t.leus@tudelft.nl). }
		{Two conferences precursors of this manuscript have been published in the Proceedings of the Twenty-Fifth European Signal Processing Conference, September 2017 \cite{Bushnaq2017EH} and the Eighteenth International Workshop on Signal Processing Advances in Wireless Communications, July 2017 \cite{Bushnaq2017dynamic}. }
		{This work was supported by the KAUST-MIT-TUD consortium grant OSR2015-Sensors-2700.}
		
	}
	
	\begin{abstract}
		The paper studies optimal sensor selection for source estimation in energy harvesting Internet of Things (IoT) networks. Specifically, the focus is on the selection of the sensor locations which minimizes the estimation error at a fusion center, and to optimally allocate power and bandwidth for each selected sensor subject to a prescribed spectral and energy budget. To do so, measurement accuracy, communication link quality, and the amount of energy harvested are all taken into account. The sensor selection is studied under both analog and digital transmission schemes from the selected sensors to the fusion center. In the digital transmission case, an information theoretic approach is used to model the transmission rate, observation quantization, and encoding. We numerically prove that with a sufficient system bandwidth, the digital system outperforms the analog system with a possibly different sensor selection.
		
		Two source models are studied in this paper: static source estimation for a vector of correlated sources and dynamic state estimation for a scalar source. The design problem of interest is a Boolean non convex optimization problem, which is solved by relaxing the Boolean constraints. We propose a randomized rounding algorithm which generalizes the existing algorithm. The proposed randomized rounding algorithm takes the joint sensor location, power and bandwidth selection into account to efficiently round the obtained relaxed solution.
	\end{abstract}
	
	
\end{frontmatter}

\section{Introduction}
Wireless sensor networks (WSN) have been gaining increasing interest in the last few years due to their role in emerging technologies such as Internet of things (IoT). Advanced sensor networks are needed in order to meet the increasing needs of IoT applications, such as automated surveillance, environmental monitoring, smart cities, and so on \cite{Zanella2014IoT,Xu2014IoT,Lin2017IoT}. To guarantee a durable autonomous sensor network, sensing nodes should be capable of processing and communicating data with restricted energy harvesting (EH) and consumption budgets. Despite the wide range of studies regarding WSN/IoT network optimization in the literature, there still are many challenges in implementing these networks. Sensors are expected to  harvest energy and control their consumption to result in self-powered sensing nodes and, on the other hand, they need to obtain accurate observations and communicate them reliably. The complexity of such problems lies in designing a mathematical model that accounts for many factors. 
Managing the available resources, the system costs and the amount of data while achieving the desired inference performance forms a major challenge in today's big sensor networks. Therefore, placing the sensors at optimal locations to gather informative data with fewer sensors and optimizing power and spectral resources is a fundamental design task.

Sensor selection (placement) is the problem of choosing the best subset of sensors (locations) from a set of candidate sensors (locations). This is a combinatorial problem, which can be solved optimally through an exhaustive search by evaluating a performance measure (e.g., inference accuracy) for all possible combinations that satisfy a budget constraint. However, this process is computationally intractable when the number of selection variables is large. Instead, a suboptimal solution can be obtained by greedily  selecting sensors one by one. Such a greedy algorithm is near optimal, if the performance measure can be expressed as a submodular set function of the selection indicators with cardinality constraints \cite{greedy2010}, \cite{greedy2015sundeep}. Alternatively, the sensor selection problem can be solved suboptimally using convex optimization \cite{boyd2009}, which utilizes the convexity of the performance measure and constraint functions to solve the optimization problem \cite{boyd2004convex},  \cite{chepuri2015sparsity}. For solutions based on convex optimization, the discrete selection variables are relaxed to the continuous domain and an approximate Boolean solution is retrieved using rounding. See~\cite{sundeep16FnT}, for an overview  on sensor selection techniques for common statistical signal processing tasks. 

Two related, yet different, major challenges in sensor networks are, (1) online sensor activation/deactivation where the sensor operation is scheduled based on real-time measurements \cite{Giannakis2012sensor_censoring, jawaid2015submodularity, savage2009optimal,wang2013sequential}. And, (2) offline sensor selection (placement) where the objective is to select a best subset of sensors (locations) out of a candidate set of sensors (locations). Offline sensor selection is done at the network design time, such that a desired performance is met based on prior statistics, which do not depend on real-time measurements \cite{greedy2010, boyd2009, chepuri2015sparsity, sparsity2016spain, zhang2015sensor, zhang2017sensor}. The focus of this paper is on the offline sensor selection.

The overall offline sensor selection is enhanced by considering different practical issues such as measurement accuracy, observation transmission quality, and EH efficiency. Considering these practical issues guarantees a better overall system performance in the sense of minimizing the minimum mean squared error (MMSE) at a central fusion center (FC). In \cite{boyd2009},\cite{chepuri2015sparsity} the sensing locations are selected based only on the measurement accuracy at the sensor level. The earlier mentioned practical considerations are addressed in \cite{sparsity2016spain}. Nevertheless, only sensing location is optimized in \cite{sparsity2016spain}, which restricts the system flexibility.  In \cite{ren2014dynamic}, the authors assume uncertainty of successful reception at a remote estimator based on the sensor transmission power and assume that the sensors receive feedback upon successful transmission. In that work, the sensor selection problem is solved such that either a low power sensor or a high power sensor is placed at each candidate sensing node. However, only two types of sensors are considered and no spectrum allocation is performed.

Our proposed estimator is carefully designed based on the measurement model.
The \textit{maximum a postiriori} (MAP) estimator reaches the MMSE given a linear measurement model for a static source (i.e. temporal correlation is ignored). While the \textit{Kalman filter} is employed instead in order to take the temporal correlation into account while achieving MMSE estimation \cite{citeulike:347166}. Modeling the unknown source while considering both the sources' cross-correlation and temporal correlation improves system optimization quality.
Despite the lack of performance guarantees, several greedy algorithms were proposed to minimize the estimation error for the vector state linear dynamical system subject to a prescribed number of sensing nodes \cite{greedy2010, summers2016actuator, zhang2015sensor, zhang2017sensor}. None of these studies has considered the cost of sending the sensor observations to the FC and the quality of the communication links between the sensors and the FC.
We focus in this paper on i) static vector source estimation and ii) dynamic scalar source estimation.

In \cite{Bushnaq2017EH, Giannakis2012sensor_censoring, boyd2009, chepuri2015sparsity, sparsity2016spain, calvo2016sensor}, a static measurement model for a vector of unknown sources was considered such that the distributed parameter estimation is minimized based on the current measurement statistics. These works considered a source without temporal correlation. In \cite{boyd2009}, the sensor placement via convex relaxation was introduced for static state estimation. For a wide range of applications, physical quantities in nature tend to change slowly over time. Hence, the temporal correlation between observations that are separated by orders of a few seconds tend to be high. Scalar state estimation is studied in \cite{Bushnaq2017dynamic, ren2014dynamic, savage2009optimal} in order to obtain a simple and optimal sensor selection solution.

Digital observation transmission is expected to perform better than analog transmission schemes because of its immunity to channel noise. Nevertheless, few studies in the literature consider digital transmission schemes in sensor selection problems because of analysis complexity. The introduced noise due to observation quantization is not Gaussian and therefore linear measurement models cannot be used directly. In \cite{Shirazi2017Digital, Shirazi2017fisher}, the sensors' total power consumption is optimized in an online fashion based on the Bayesian Fisher information at the FC which receives quantized sensor observations. However, the amount of energy available through EH at each sensor is not taken into account. 
 
 In this work, practical aspects such as the sensor's EH and observation transmission quality to the FC are taken into account. The main goal of this paper is to combine optimal sensor placement with novel and important selection dimensions that add to the network design flexibility, namely, we allow for transmission power and resource block (i.e., time-frequency channel) allocation. The sensor's transmission power is optimized by considering different kinds of sensors where expensive sensors are supplied with more EH capabilities and higher battery capacities as compared to cheaper sensors. In this setup, the approach in \cite{ren2014dynamic} is generalized by considering $K$ sensor kinds. Further, we allow sensors to transmit their observations over different spectral bandwidths such that the total system bandwidth is limited. In spite of the general awareness of the energy scarcity in IoT networks, only few studies allocate the sensor transmission power level for dynamic estimation. We present sensor selection solutions considering analog and digital transmission schemes and compare their performances. In the digital scheme, we analyze observation quantization and encoding based on an information theoretic approach. The number of quantization levels is optimized based on the allocated bandwidth and the signal to noise ratio between the candidate sensor and the FC. We obtain a suboptimal sensor selection via convex optimization by relaxing the discrete variables and rounding the obtained solution. A novel rounding algorithm is proposed in order to enhance the rounding efficiency. The contribution of this paper can be summarized as follows:
\begin{itemize}
	\item The sensor transmission power and operating bandwidth are jointly optimized with the sensing location. This gives network designers the flexibility to place more expensive sensors with a higher power budget and data rate in strategic locations while cheaper sensors are placed in less important locations. 
	\item We model a practical system which takes the EH, channel gain and measurement accuracy into account. Similar considerations were taken into account in \cite{Bushnaq2017EH, Bushnaq2017dynamic, calvo2016sensor, sparsity2016spain}, however, all of them assumed analog communication where sensors directly amplify and forward observations.
	\item Sensor selection is optimized for digitally transmitted observations to the FC. An information theoretic approach is utilized to express the quantization and channel error.
	\item A generalized randomized rounding algorithm is proposed in order to efficiently round the relaxed solutions taking the joint power, location and resource block selection into account.
\end{itemize}

\textbf{\textit{Notation:}} Throughout the paper, lower-case letters  $x$  denote variables, while boldface lower-case letters  ${\bf x}$ and boldface upper-case letters ${\bf X}$  denote vectors and matrices, respectively. ${\bf 1}_x$ and ${\bf I}_x$ denote the ones vector of size $x$ and the identity matrix of size $x \times x$, respectively. $ \{\cdot \}^T  $ denotes the transpose operator. The operator $\mathbb{E}\{\cdot\}$ denotes expectation. For a vector ${\bf x}$, the operator $||{\bf x}||_p$ denotes the $\ell_p$ norm. For matrix ${\bf X}$, the operator $tr({\bf X})$ denotes the trace operator. The calligraphic font $\mathcal{X}$ refers to sets. The floor function is denoted by $\floor{\cdot}$. Finally, $\mathbb{R}$ and $\mathbb{Z}$ denote the sets of real and integer numbers, respectively.

The paper is organized as follows. In the next section, the system setup and the problem statements are discussed. In Sections \ref{static} and \ref{dynamic}, the sensor selection problem is formulated and solved for the static source and the dynamic source, respectively.  Then, the randomized rounding algorithm is discussed in Section \ref{rounding_algorithms}. Numerical experiments are presented in Section \ref{num_experiments} before we conclude this work.

\section{System setup and problem statement}

\begin{figure}
\centering
\begin{tikzpicture}
\draw (0, 0) node[inner sep=0] {\includegraphics[width=9cm]{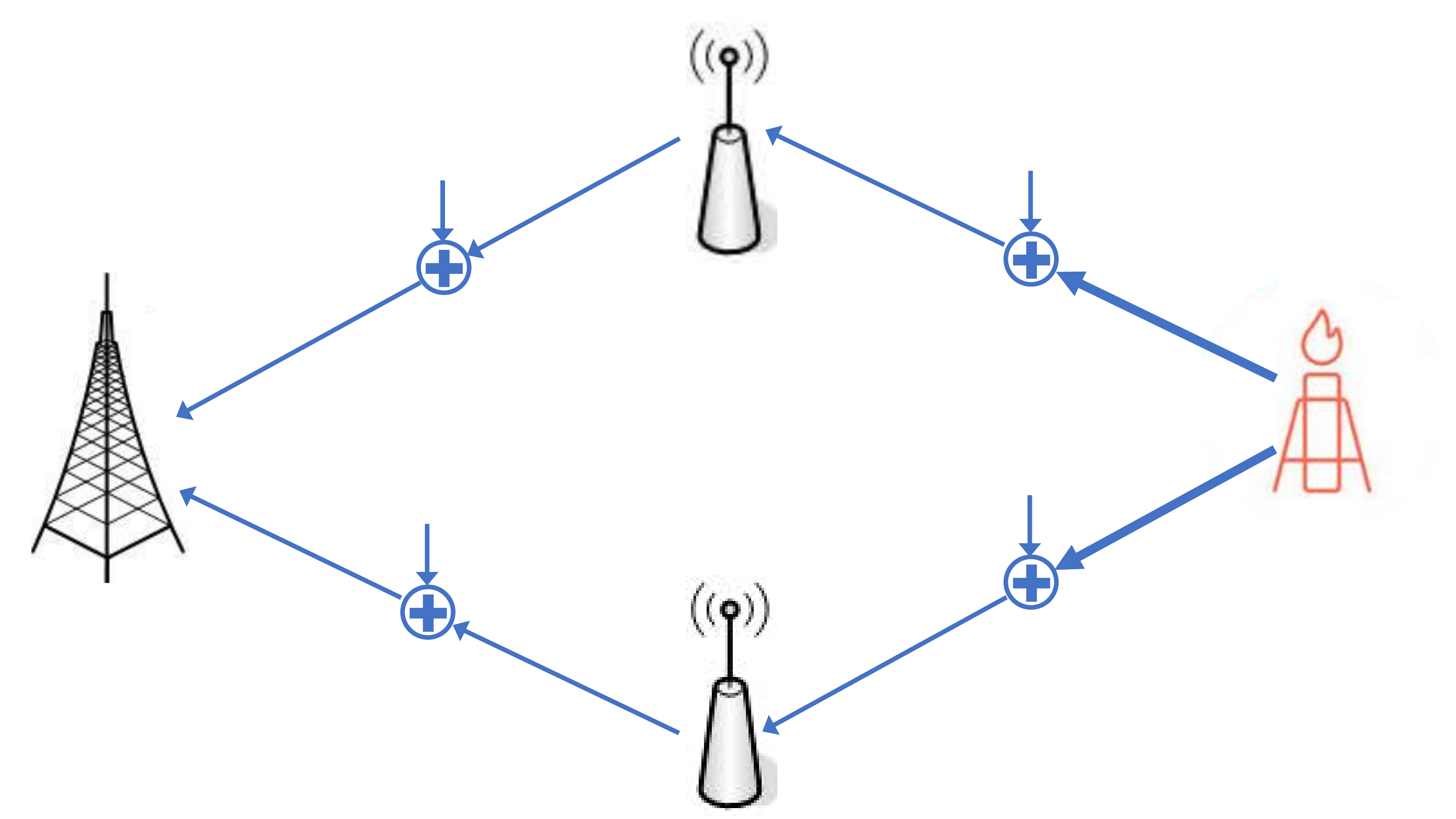}};
\draw (-1.8, 1.6) node {$\phi_1 $};
\draw (-1.8, -.4) node {$\phi_L $};
\draw (1.85, 1.6) node {$v_1 $};
\draw (1.85, -.4) node {$v_L $};
\draw (-.8, 1.6) node {$g_1$};
\draw (-.8, -1.4) node {$g_L$};
\draw (2.7, .75) node {${\bf h}_1$};
\draw (2.7, -.3) node {${\bf h}_L$};
\draw (2.7, 0.2) node {\vdots};
\draw (-3, -.1) node {\vdots};
\draw (0, -2.65) node {Sensors};
\draw (3.7, .8) node {${\boldsymbol{\theta}}$};
\draw (3.7, -.65) node {Source};
\draw (0, -2.65) node {Sensors};
\draw (-3.8, -1.2) node {FC};
\end{tikzpicture}
		\caption{System setup.}
		\label{fig:channels}

\end{figure}
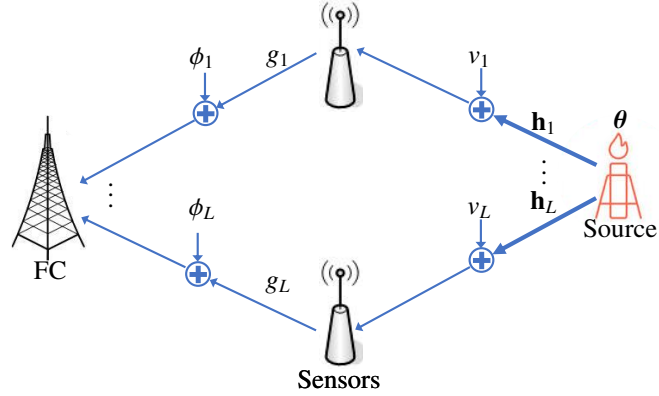

Consider estimating a vector of unknown sources $\boldsymbol{ \theta} \in \mathbb{R}^m$ which is assumed to be a zero-mean Gaussian random vector with covariance matrix ${\boldsymbol \Sigma}_{{\boldsymbol \theta}}$, i.e., ${\boldsymbol{\theta}} \sim \mathcal{N} ({\bf 0},{\boldsymbol{ \Sigma}_{\boldsymbol{ \theta}}})$. We can place sensors at a subset of predefined candidate sensing locations ${\mathcal P} = \{ {\bf p}_1, \dots, {\bf p}_L \} $ to measure the unknown source parameters. The deployed sensors send their observations over a limited system bandwidth of $W $ [Hz] to a FC (more specifically, each sensor shares a part of the available $W$ [Hz]), where the collected information is utilized to estimate the vector of unknown source parameters. Figure \ref{fig:channels} illustrates the system setup. To create an autonomous system, the deployed sensors are equipped with energy harvesting (EH) capabilities. We consider that there are different sensor types $ {\mathcal T} = \{ t_0, \dots, t_K \}$, where different sensor types measure the same quantities with the same measurement accuracy but differ in their EH efficiency, $\eta_k$, battery capacity, $\varepsilon_k$ and cost, $c_k$. For example, more expensive sensor types are equipped with more efficient EH capabilities. The type $t_0$ is an auxiliary type with $\varepsilon_0=\eta_0=c_0=0$ representing no sensor placement\footnote{Throughout the paper, we express no sensor placement at $ {\bf p}_l$ as selecting a sensor of type $t_0$ at that location.}. Sensors send their observations over one of $B$ available transmission bandwidths, $\mathcal{W} = \{ w_1, \cdots, w_B \} $. The objective is to select the type of sensor and the bandwidth at each candidate sensing location such that the system performance is optimized. Further details about the system model are provided next.


\subsection{Measurement Modeling}
Consider a linear measurement model. The observation at the sensor placed at ${\bf p}_l$ is given by,
\begin{align} \label{measurement_model}
x_l[t] &= {\bf h}_l^T {\boldsymbol \theta}[t]+v_l[t] \\
{\boldsymbol{ \theta}}[t] &= {\bf A} {\boldsymbol{ \theta}}[t-1] +{\bf u}[t] 
\end{align}
where, ${\bf h}_l \in \mathbb{R}^m$ is the regressor (also called gain) and $v_l$ is zero-mean Gaussian noise with variance $\sigma_v^2$ that is independent from the observations at other sensors. The matrix ${\bf A} \in \mathbb{R}^{m\times m}$ is the state transition matrix and ${\bf u} \in \mathbb{R}^m$ is the driving or excitation noise. 

This model allows for accurate estimator design since both the cross-correlation between source parameters and the temporal correlation of the source parameters are taken into account. 
We focus in this paper on two special cases:
\begin{itemize}
	\item Static vector source estimation where only the correlation between the different parameters is exploited but the temporal correlation is ignored. In this case the measurement model simplifies to 
	\begin{align} \label{measurement_model_static}
	x_l &= {\bf h}_l^T {\boldsymbol \theta}+v_l.
	\end{align}
	The dependence on time, $[t]$, is removed since the estimation at each time slot may be performed independently. 
	\item Dynamic scalar source estimation where the correlation between the different parameters is ignored but temporal correlation is exploited. The measurement model in this case is rewritten as,
	\begin{align} \label{measurement_model_dynamic}
	x_l[t] &= { h}_l { \theta}[t]+v_l[t], \\
	{{ \theta}}[t] &= {a} {{ \theta}}[t-1] +{u}[t]. 
	\end{align}
\end{itemize}

\subsection{Resource Block Allocation}
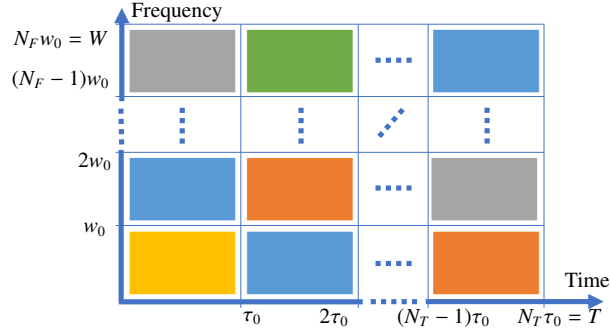
\begin{figure}
	\centering
	\begin{tikzpicture}[thick,scale=0.9, every node/.style={transform shape}]
	\draw (0, 0) node[inner sep=0] {\includegraphics[width=9cm]{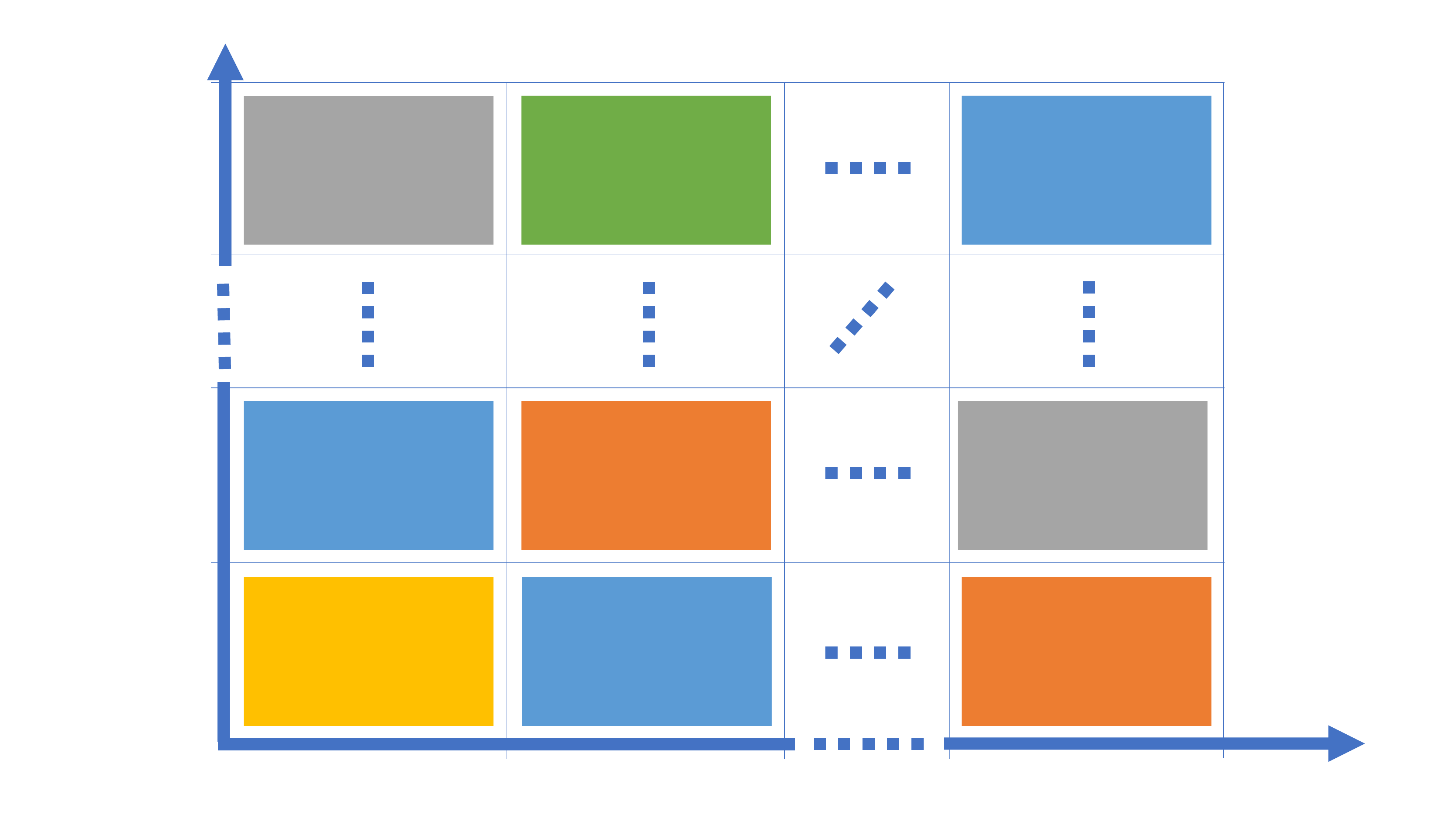}};
	\draw (-4, 1.8) node {\small $N_F w_0=W $};
	\draw (-4, 1.2) node {\small$(N_F-1) w_0 $};
	\draw (-3.5, 0) node {\small$ 2w_0 $};
	\draw (-3.5, -1.0) node {\small$ w_0 $};	
	\draw (-1.2, -2.3) node {\small$ \tau_0 $};
	\draw (-.0, -2.3) node {\small$ 2\tau_0 $};
	\draw (1.6, -2.3) node {\small$ (N_T-1)\tau_0 $};
	\draw (3.3, -2.3) node {\small$ N_T\tau_0=T $};
	\draw (3.7, -1.75) node {\small Time};
	\draw (-2.3, 2.2) node {\small Frequency};
	\end{tikzpicture}
	\caption{Time frequency channels. Each selected sensor (represented by a color) shares a subset of the channels.}\label{fig:tx_schedul}
\end{figure}

The communication channels between the sensors and the FC are assumed orthogonal (i.e. no interference between channels). To validate this assumption, the transmission is scheduled over time (TDMA) and/or frequency (FDMA). Let sensors transmit one observation every fixed transmission interval of $T$ [s] over a total system bandwidth of $W$ [Hz]. As shown in Figure \ref{fig:tx_schedul}, the transmission interval and available bandwidth are divided into $N = N_T N_F$ channels, where $N_T$ and $N_F$ are the number of time and frequency channels, respectively. Denoting the channel interval as $\tau_0 =\frac{T}{N_T}$, the channel bandwidth, $w_0 = \frac{W}{N_F}$ reduces to $w_0 = \varsigma \frac{1}{\tau_0}$ where $\varsigma$ accounts for the modulation and pulse shaping schemes. Without loss of generality, we assume that $\varsigma = 1$.

The $N$ channels are shared among the selected sensors such that each sensor can transmit over one of $B$ predefined number of channels, $\mathcal{N} = \{ {N}_1, \cdots, {N}_B \} $. Based on the number of channels ${N}_b$ given to a sensor, we define a {\it resource block} as, 
\begin{align}
w_b = \frac{W {N}_b}{N} = w_0 \dfrac{N_b}{N_T},
\end{align}
where the total number of channels cannot exceed $N$. The resource block is the total bandwidth used by a sensor multiplied by the percentage of time resource occupation over the that bandwidth.\footnote{The term {\it 'resource block allocation'}, is interchanged with {\it 'bandwidth allocation'} throughout the paper.}

\underline{Example:} Let $T = 1 [s]$ and $W=1 M[Hz]$ be divided into $N_T =10$ and $N_F=100$ time and frequency channels, respectively. Therefore, each transmission channel has $\tau_0 = 100 m[s]$ and $w_0=10K[Hz]$. The number of channels is $N=1000$ channels which are shared by all selected sensors. For $\mathcal{N} = \{10,20,50\}$, the resource block allocation set is $\mathcal{W} = \{ 10K,20K,50K\} [Hz]$.

\subsection{Power Allocation}

\begin{figure}
	\begin{center}
		\includegraphics[width=7.5cm]{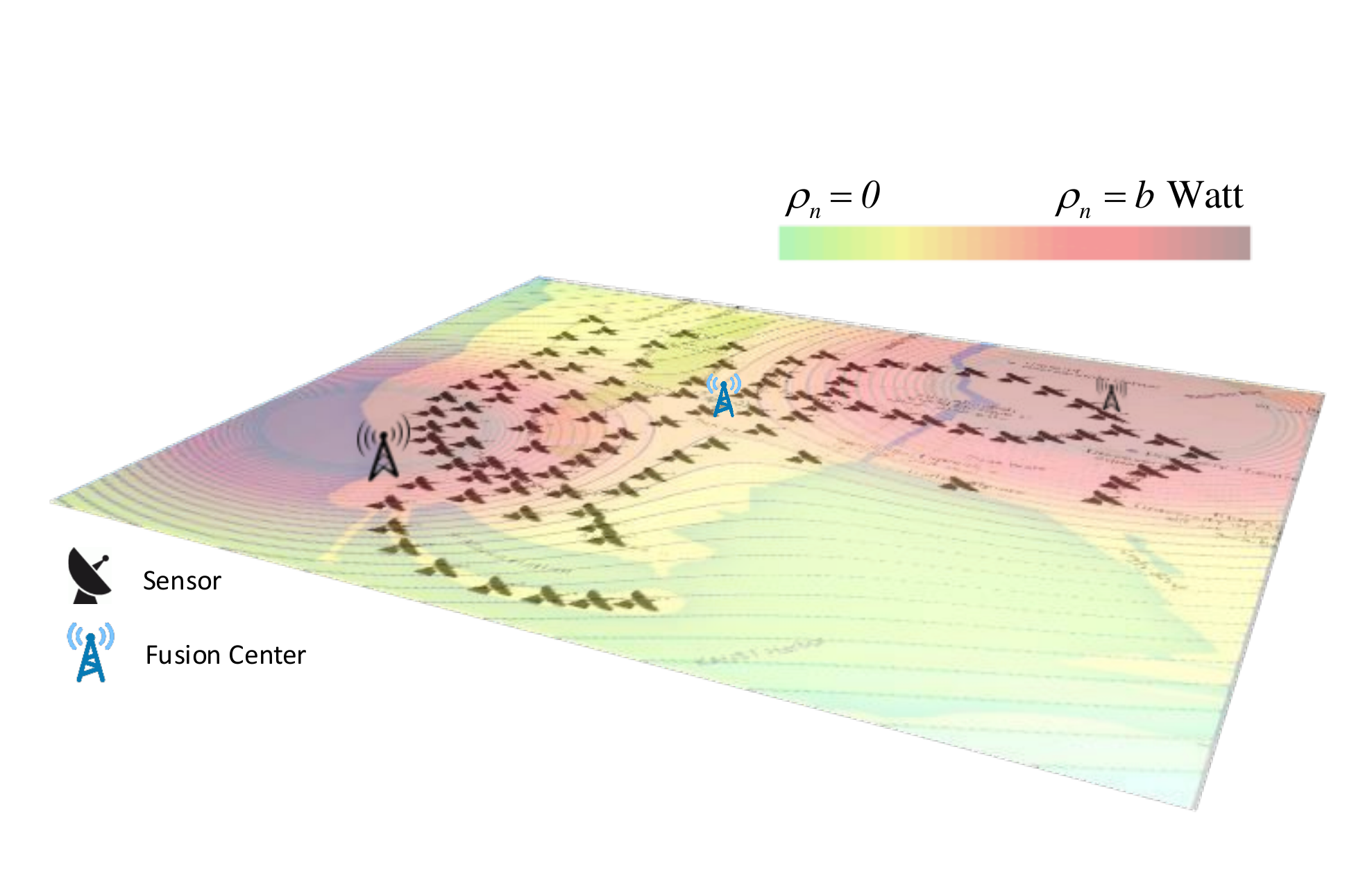}
		\caption{ Candidate sensing locations, fusion center and EH distribution, $\rho_{l}$ over the area of interest.}
		\label{fig:map}
	\end{center}
\end{figure}

The sensor type selection is equivalent to discrete power allocation. Since we assume that the EH amount is location dependent, the selected sensor transmission power is a function of the available energy at ${\bf p}_l$ as well as the energy harvesting efficiency of the deployed sensor type $t_k$. To be more specific, the transmission power will be $P_{l,k} = f(\rho_{l},\eta_k, \varepsilon_k)$, where $\rho_l$ is the average power available at ${\bf p}_l$. For instance, the transmission power can be formulated as, $P_{l,k} = \min(\rho_{l}\eta_k, \varepsilon_k)$, where $\varepsilon_k$ is a positive constant representing an upper limit for EH, e.g., battery capacity. Figure \ref{fig:map} shows an illustration of the average EH intensity, $\rho_l$, over candidate sensor locations.


\subsection{Channel Modeling}

The sensor located at ${\bf p}_l$ transmits its observation with power $P_{l,k}$ to the FC over a deterministic AWGN channel with channel gain $g_l$ and receiver noise $\phi_l \sim \mathcal{N}(0,\sigma_{\phi}^2)$. The channel gain is given as $g_l = d^{-\alpha}{({\bf p}_l,{\bf p}_{\rm FC})} $ where $d{({\bf p}_l,{\bf p}_{\rm FC})}$ is the distance between the sensing location ${\bf p}_l$ and the location of the FC, ${\bf p}_{\rm FC}$, and $\alpha$ is the path loss exponent. It is assumed that ${\bf h}_l$, $\boldsymbol{ \Sigma}_{\theta}$, $g_l$ and $\sigma_{\phi}^2$ are known at the FC. It is also assumed that the FC has the statistics of the average EH over time at each sensor location, i.e., $P_{l,k}$ is known. Figure \ref{fig:channels} summarizes the system setup.

Before discussing the problem statement, we formally express the signal to noise ratio (SNR) per channel use at the FC in the following proposition.

{\bf Proposition 1:} Given a system bandwidth of $W$ [Hz], transmission interval $T$ [s], and sensor transmission power, $P_{l,k}$, the signal to noise ratio (SNR) per channel use at the FC is independent of $N_F$ and $N_T$ and is inversely proportional to the sensor bandwidth allocation,
\begin{align}\label{eq:SNR}
{\rm{SNR}}_{l,k,b} =   \frac{P_{l,k} g_l  }{ \kappa \Delta w_b} ,
\end{align}

{\bf Proof:} Assuming that the amount of energy a sensor collects over $T$ seconds, $P_{l,k}$, is divided over the the number  of channels the sensor uses to transmit its observation. Therefore, the amount of energy per channel is,
\begin{align}\label{eq:tx_energy_per_n}
\hat{E}_{l,k,b} = \frac{P_{l,k} T}{{N}_b}.
\end{align}
Given that the channel interval is equal to $\tau_0$, the transmission power per channel is expressed as,
\begin{align}\label{eq:tx_power_per_n}
\hat{P}_{l,k,b} = \frac{P_{l,k} T}{\tau_0 {N}_b}.
\end{align}
 The receiver noise power $\sigma_{\phi}^2$, is a function of the channel bandwidth, i.e., 
\begin{equation}\label{eq:reciever_noise}
\sigma_{\phi}^2 = \kappa \Delta  w_0,
\end{equation}
where, $\kappa \approx 1.3807 \times 10^{-23}$ Joule per Kelvin is Boltzmann's constant, $\Delta$ is the receiver absolute temperature and $w_0$ is the transmission bandwidth.

By combining \eqref{eq:tx_power_per_n} and \eqref{eq:reciever_noise}, and considering the channel gain, the $\rm{SNR}$ at the FC is expressed as,
\begin{align}\label{eq:SNR_proof}
{\rm{SNR}}_{l,k,b} =  \frac{P_{l,k} g_l  T}{ \kappa \Delta w_0  \tau_0 {N}_b} = \frac{P_{l,k} g_l  }{ \kappa \Delta } \dfrac{N}{{N}_b W} =  \frac{P_{l,k} g_l  }{ \kappa \Delta w_b} .
\end{align}

\subsection{Problem Statement}
Let $ \mathcal{S}_{(l,k,b)}$ represents the sensor located at ${\bf p}_l$, $\forall l \in \{1,\cdots, L\}$, with the type $t_k$, $\forall k \in \{0,\cdots, K\}$ and transmission bandwidth $w_b$ [Hz], $\forall b \in \{1,\cdots, B\}$. The objective is then to select a subset of $\mathcal{S} = \{\mathcal{S}_{(l,k,b)} \, | \, \forall \{l,k,b\}\}$ such that the estimator of the source parameters, $\hat{\boldsymbol\theta}$, at the FC is as close as possible in terms of MMSE to the original value, $\boldsymbol\theta$, subject to system cost and bandwidth constraints. Equivalently, we can minimize the system cost/bandwidth subject to an upper bound on the estimation error. In the following two sections, we consider achieving these goals for a static vector source and a dynamic scalar source, respectively.

\section{Static Source}
\label{static}

The static source model is used when the source parameters do not change over time. We study in this section the sensor selection for a static source considering two schemes of sensor observation transmission to the FC, namely, analog and digital transmission schemes.

\subsection{Analog Transmission Scheme}
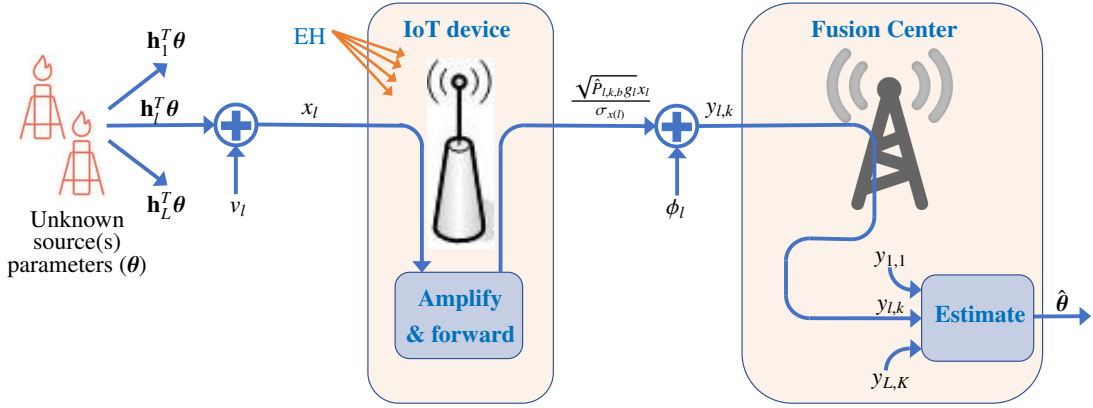
\begin{figure*}
	\centering
	\begin{tikzpicture}[thick,scale=0.9, every node/.style={transform shape}]
	\draw (0, 0) node[inner sep=0] {\includegraphics[width=16cm]{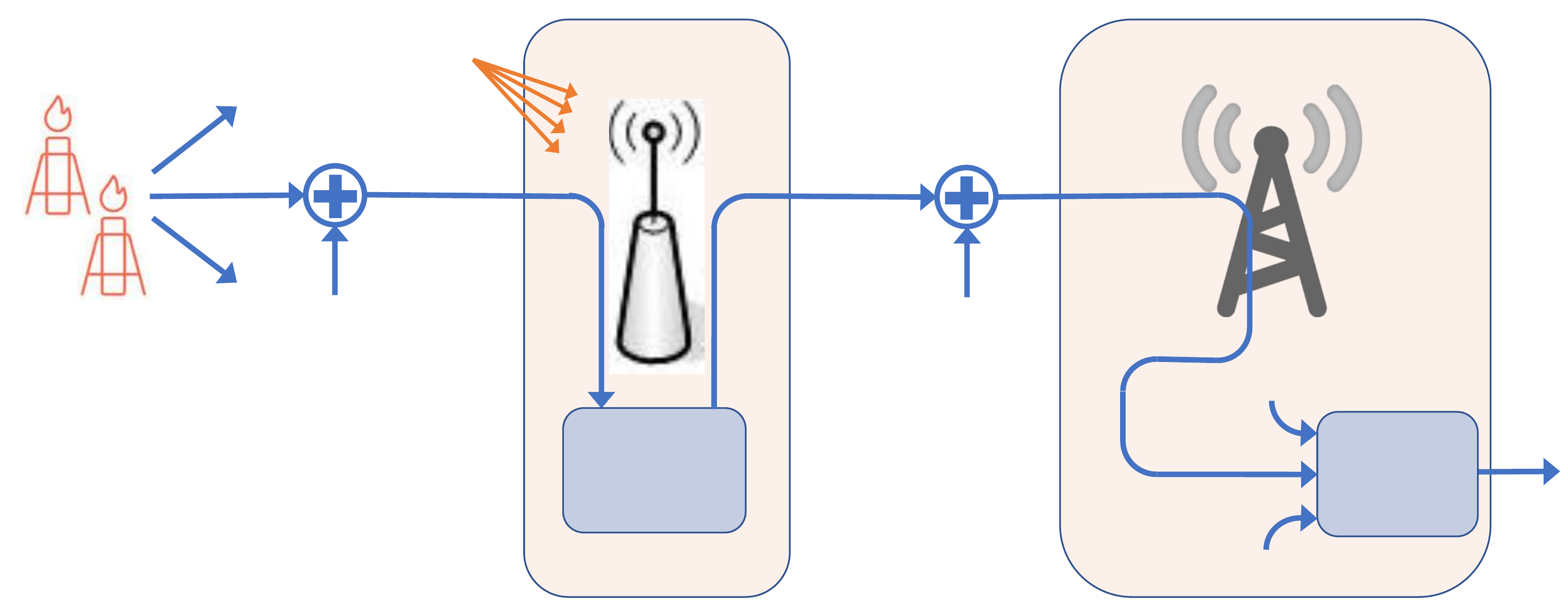}};
	\draw (-6.9, -.3) node { Unknown };
	\draw (-6.9, -.65) node {source(s) };
	\draw (-6.9, -1) node {parameters (${\boldsymbol{ \theta}}$)};
	\draw (-5.6, 2.3) node {${\bf h}_1^T {\boldsymbol{ \theta}}$};
	\draw (-5.7, 1.3) node {${\bf h}_l^T {\boldsymbol{ \theta}}$};
	\draw (-5.6, -.1) node {${\bf h}_L^T {\boldsymbol{ \theta}}$};
	\draw (-4.55, -.15) node {$v_l $};
	\draw (-3.5, 1.3) node {$x_l $};
	
	\draw (-3.5, 2.4) node {{\color{RoyalBlue}EH}};
	\draw (-1.35, 2.5) node {{\color{RoyalBlue} {\bf IoT device}}};
	\draw (-1.3, -1.5) node {{\color{RoyalBlue} {\bf Amplify}}};	
	\draw (-1.3, -2) node {{\color{RoyalBlue} {\bf \& forward}}};
	
	\draw (.9, 1.5) node {$\frac{\sqrt{\hat{P}_{l,k,b}g_l} x_l}{\sigma_{x(l)} } $};
	\draw (1.85, -.15) node {$\phi_l $};
	\draw (2.5, 1.3) node {$y_{l,k} $};
	
	\draw (4.9, 2.5) node {{\color{RoyalBlue} {\bf Fusion Center}}};
	\draw (6.3, -1.7) node {{\color{RoyalBlue} {\bf  Estimate}}};
	\draw (5.0, -.9) node {$y_{1,1}$};
	\draw (5.0, -1.6) node {$y_{l,k}$};
	\draw (5.0, -2.7) node {$y_{L,K}$};
	\draw (7.5, -1.5) node {$\hat{\boldsymbol \theta}$};
	\end{tikzpicture}
	\caption{System Model. The IoT device includes the sensor, the EH equipment and the wireless transmission system at ${\bf p}_l$.}\label{fig:sys_mdl_A}
\end{figure*}

Recall that the observations at the sensor level are as given in \eqref{measurement_model_static}. The selected sensor at ${\bf p}_l$ amplifies its observation based on the available power and bandwidth, ${\hat P}_{l,k,b}$ and forwards it to the FC. The analog system model is described in Figure \ref{fig:sys_mdl_A}.
At the FC, the received signal from $\mathcal{S}_{(l,k,b)}$ is expressed as, 
\begin{align}\label{eq:y}
y_{l,k,b} = s_{l,k,b} \bigg(\frac{\sqrt{\hat{P}_{l,k,b}g_l} x_l}{\sigma_{x(l)} }  + \phi_l \bigg).
\end{align} 	
where $s_{l,k,b} $ is a selection indicator with $s_{l,k,b}=1 $ indicating the selection of $\mathcal{S}_{(l,k,b)}$ and  $s_{l,k,b} =0$ indicating otherwise. We assume that $\phi_l$, $v_l$ and $\boldsymbol{ \theta}$ are uncorrelated. To force the average transmitted power to $\hat{P}_{l,k,b}$, the transmission signal is scaled by $\sigma_{x(l)}$, where $\sigma_{x(l)}^2$ denotes the average power of the measurement $x_l$ and is given by
\begin{eqnarray}\label{eq:var_x}
\sigma_{x(l)}^2 = \mathbb{E}\{|x_l|^2\} = \mathbb{E}\{|{\bf h}_l^T {\boldsymbol\theta} +v_l|^2\} = {\bf h}_l^T {\boldsymbol\Sigma}_{\boldsymbol{\theta}}{\bf h}_l +\sigma_{v}^2.
\end{eqnarray}
Note that $\sigma_{x(l)}^2$ is assumed to be known at the sensor. Since the estimation error covariance matrix is a function of the received signal ${\rm SNR}$, \eqref{eq:y} can be normalized as \cite{kay1993fundamentals}
\begin{eqnarray}\label{eq:received_f_c}
 y_{l,k,b} &=& s_{l,k,b} ({\bf h}_l^T {\boldsymbol \theta} + e_{l,k,b} ),
\end{eqnarray} 	
 where, $e_{l,k,b} = v_l +\dfrac{\phi_l \sigma_{x(l)}}{\sqrt{\hat{P}_{l,k,b} g_l}} $ is the equivalent noise. Note that $e_{l,k,b}$ is a zero-mean Gaussian noise with variance
 \begin{eqnarray}\label{eq:analog_noise_variance}
 \sigma_{e(l,k,b)}^2 = \sigma_v^2 + \dfrac{({\bf h}_l^T {\boldsymbol\Sigma_{\theta}} {\bf h}_l + \sigma_v^2)\sigma_{\phi}^2}{g_l  \hat{P}_{l,k,b}}.
 \end{eqnarray}	
  $\sigma_{e(l,k,b)}^2$ is the aggregate noise variance of the observation and receiver noises.

Based on the observations received at the FC given by \eqref{eq:received_f_c}, the unknown parameters can be reconstructed using the MMSE estimator. Denoting the MMSE estimate of ${\boldsymbol \theta}$ as $\hat{\boldsymbol \theta}$, the MMSE error covariance matrix, $ \boldsymbol{\Sigma}_{\boldsymbol{\theta}|\bf{y}} = \mathbb{E}\{ ( {\boldsymbol \theta} - \hat{\boldsymbol \theta})( {\boldsymbol \theta} - \hat{\boldsymbol \theta})^T  \} $ is expressed as, \cite{kay1993fundamentals}

\begin{equation}\label{eq:MSE_matrix}
\boldsymbol{ \Sigma}_{{\boldsymbol\theta}|{\bf y}}({\bf S}) =\left( {\boldsymbol \Sigma}_\theta^{-1}+\sum\limits_{l =1}^{L} \sum\limits_{k =1}^{K} \sum\limits_{b =1}^{B} \dfrac{ s_{l,k,b}}{ \sigma_{e(l,k,b)}^2    }   {\bf h}_l {\bf h}_l^T   \right)^{-1},
\end{equation}	
where {$\bf y$} encompasses all the received observations at the FC. Observe how the selection indicators \{$s_{l,k,b}$\} in the numerator affect the MMSE error covariance matrix. All the selection indicators are encompassed in the set of matrices ${\bf S} = \{ {\bf S}_1, \cdots, {\bf S}_L \}$ where ${\bf S}_l$ includes the selection indicators for all the sensor type and bandwidth combinations at the sensing location ${\bf p}_l$ as follows,
\[ {\bf S}_l=
\begin{bmatrix}
s_{l,0,1}      & s_{l,0,2}      & \dots          &  s_{l,0,B} \\
s_{l,1,1}      & s_{l,1,2}          & \dots       &  s_{l,1,B} \\
\vdots       & \vdots                & \ddots      &  \vdots \\
s_{l,K,1}      & s_{l,K,2}          & \dots       &  s_{l,K,B} \\
\end{bmatrix},\]
where the element at the $(k+1)$-th row and the $b$-th column is set to one if the sensor $\mathcal{S}_{l,k,b}$ is selected. 

{\bf Proposition 2:}
Increasing the allocated bandwidth and sending the same copy of a sensor's observation over multiple transmission channels does not improve the estimation performance for the \textit{analog transmission scheme}, given a fixed transmission energy per observation. 

{\bf Proof:} See Appendix \ref{app:bandwidth_increase_analog}.

Based on Proposition 2, we can let each sensor transmit over one channel to save bandwidth, i.e., $N_b = 1$ and $w_b=\frac{W}{N} =\frac{w_0}{N_T}$. Consequently, $\hat{P}_{l,k,b}$, $\sigma_{e(l,k,b)}^2$ and $s_{l,k,b}$ are reduced to $\hat{P}_{l,k}$, $\sigma_{e(l,k)}^2$ and $s_{l,k}$.

The reconstruction error is a function of the error covariance matrix. To guarantee a small reconstruction error, one might, for example, minimize the sum of the eigenvalues of the error covariance matrix (known as the A-optimality criterion), denoted by
\begin{equation}\label{eq:error_tr}
tr \{	\boldsymbol{\Sigma}_{{\boldsymbol\theta}|{\bf y}}({\bf s}) \} = tr\left( {\boldsymbol \Sigma_{\boldsymbol \theta}}^{-1} + \sum\limits_{l =1}^{L} \sum\limits_{k =0}^{K} \dfrac{s_{l,k}}{\sigma_{e(l,k)}^2 }{\bf h}_l {\bf h}_l^T  \right)^{-1},
\end{equation}
where ${\bf s} = \{{\bf s}_1, \cdots, {\bf s}_L \}$ is defined as a set of vectors with ${\bf s}_l$ indicating the sensor type at the location ${\bf p}_l$. Recall that the element $s_{l,k}$ is equal to 1 if the sensor at location ${\bf p}_l$ and type $t_k$ is selected, otherwise, $s_{l,k} =0$. We assume that no more than one sensor can be selected at any location. Therefore, the $\ell_0$ norm of the vector including all sensor types at location ${\bf p}_l$, ${\bf s}_l =[s_{l,0}, \cdots, s_{l,K}]^T$, is equal to one. Having the auxiliary sensor type that represents no sensor selection, $t_0$, with $\varepsilon_0 = \eta_0 = c_0 =0$, the relation $ || {\bf s}_l||_0 = 1 $ holds whether a sensor at ${\bf p}_l$ is selected or not.

Given $L$ candidate sensing locations and $K$ sensor types with different EH capabilities, battery capacities and prices, we would like to jointly find the optimal sensor {\textit{location and power selection for MMSE-based static source estimation (Static Source LoPS)}} subject to constraints on the cost and bandwidth. The MMSE estimation error is caused by the noisy measurements and the noisy communication channels between the sensors and the FC. Since each sensor transmits over only one channel, we assume that the bandwidth constraint is always satisfied. The \textit{Static Source LoPS} optimization problem can now be formulated as

{\bf Problem 1: \textit{ Static Source LoPS:}}
\begin{align}
\underset{\{s_{l,k}\}}{\arg\min}
& \quad  tr \{	\boldsymbol{\Sigma}_{{\boldsymbol\theta}|{\bf y}}({\bf s}) \}
\label{eq:non_convex_1}\\
\text{subject to}
& \quad  \sum \limits_{l=1}^{L}{\bf c}^T {\bf s}_l \leq {\lambda} 
\tag{\ref{eq:non_convex_1}a} \label{eq:non_convex_1_a}\\
& \quad  \sum \limits_{l=1}^{L} [0 \; {\bf 1}_{K}^T ] {\bf s}_l \leq N
\tag{\ref{eq:non_convex_1}b} \label{eq:non_convex_1_b}\\
& \quad s_{l,k}  \in \{0,1\}, \quad {\forall l, k}
\tag{\ref{eq:non_convex_1}c} \label{eq:non_convex_1_c}\\
& \quad || {\bf s}_l||_0 = 1, \quad {\forall l},
\tag{\ref{eq:non_convex_1}d} \label{eq:non_convex_1_d}
\end{align}
where $\lambda$ is a prescribed system cost and ${\bf c} = [c_0 \, \dots c_K]^T$ is the cost vector for all the sensor types. The constraint \eqref{eq:non_convex_1_a} is to limit the total deployed sensor cost to $\lambda$ and the constraint \eqref{eq:non_convex_1_b} is to limit the system bandwidth to $W$ [Hz] by limiting the total number of channels used by all deployed sensors to $N$. Note how the selection of the sensor type $t_0$ does not add to the LHS of \eqref{eq:non_convex_1_b}. The constraints \eqref{eq:non_convex_1_c} and \eqref{eq:non_convex_1_d} guarantee that the selection indicators are either zero or one and that at most one sensor is deployed at each sensing location. Alternatively, the system cost can be minimized subject to a prescribed reconstruction error, $\xi$, which may be beneficial for applications in which the goal is to minimize the system cost, i.e.,
\begin{align}
\underset{\{s_{l,k}\}}{\arg\min}
& \quad \sum \limits_{l=1}^{L}{\bf c}^T {\bf s}_l
\label{eq:non_convex_2}\\
\text{subject to}
& \quad  tr \{	\boldsymbol{\Sigma}_{{\boldsymbol\theta}|{\bf y}}({\bf s}) \} \leq \xi 
\tag{\ref{eq:non_convex_2}a} \label{eq:non_convex_2_a}\\
& \quad \text{constraints }\eqref{eq:non_convex_1_b}, \eqref{eq:non_convex_1_c} \text{ and } \eqref{eq:non_convex_1_d}.
\tag{\ref{eq:non_convex_2}b} \label{eq:non_convex_2_b}
\end{align}

Although the function $tr \{\boldsymbol{\Sigma}_{{\boldsymbol\theta}|{\bf y}}({\bf s}) \} $ is convex over $ {\bf s} \in \mathbb{R}^{K+1}$ \cite{chepuri2015sparsity}, the optimization problems \eqref{eq:non_convex_1} and \eqref{eq:non_convex_2} are not convex because of the non-convex Boolean constraints in \eqref{eq:non_convex_1_c} and the $\ell_0$ norm constraints in \eqref{eq:non_convex_1_d}. To obtain a convex problem which can be solved using well-established tools, the constraints \eqref{eq:non_convex_1_c} are relaxed to $s_{l,k} \in [0,1], \ {\forall l, k}$ and the constraints \eqref{eq:non_convex_1_d} are relaxed to {$ {\bf 1}^T {\bf s}_l =1, \ {\forall l}$}. The convex relaxation of \eqref{eq:non_convex_1} can then be written as, 
\begin{align}
\underset{\{s_{l,k}\}}{\arg\min}
& \quad  tr \{	\boldsymbol{\Sigma}_{{\boldsymbol\theta}|{\bf y}}({\bf s}) \} 
\label{eq:convex_2}\\
 \text{subject to}
 & \quad \text{ constraints } \eqref{eq:non_convex_1_a} \text{ and } \eqref{eq:non_convex_1_b} 
 \tag{\ref{eq:convex_2}a} \label{eq:convex_2_a}\\
& \quad s_{l,k}  \in [0,1], \quad {\forall l,k}
\tag{\ref{eq:convex_2}b} \label{eq:convex_2_b}\\
& \quad {\bf 1}^T {\bf s}_l =1, \quad {\forall l}.
\tag{\ref{eq:convex_2}c} \label{eq:convex_2_c}
\end{align}

The solution of \eqref{eq:convex_2} will be between zero and one. Hence, a rounding heuristic should be applied to the solution to obtain a Boolean solution \cite{chepuri2015sparsity,boyd2009}. These heuristics are discussed in Section \ref{rounding_algorithms}.

\subsection{Digital Transmission Scheme} 
\begin{figure*}
	\centering
	\begin{tikzpicture}
	\draw (0, 0) node[inner sep=0] {\includegraphics[width=16cm]{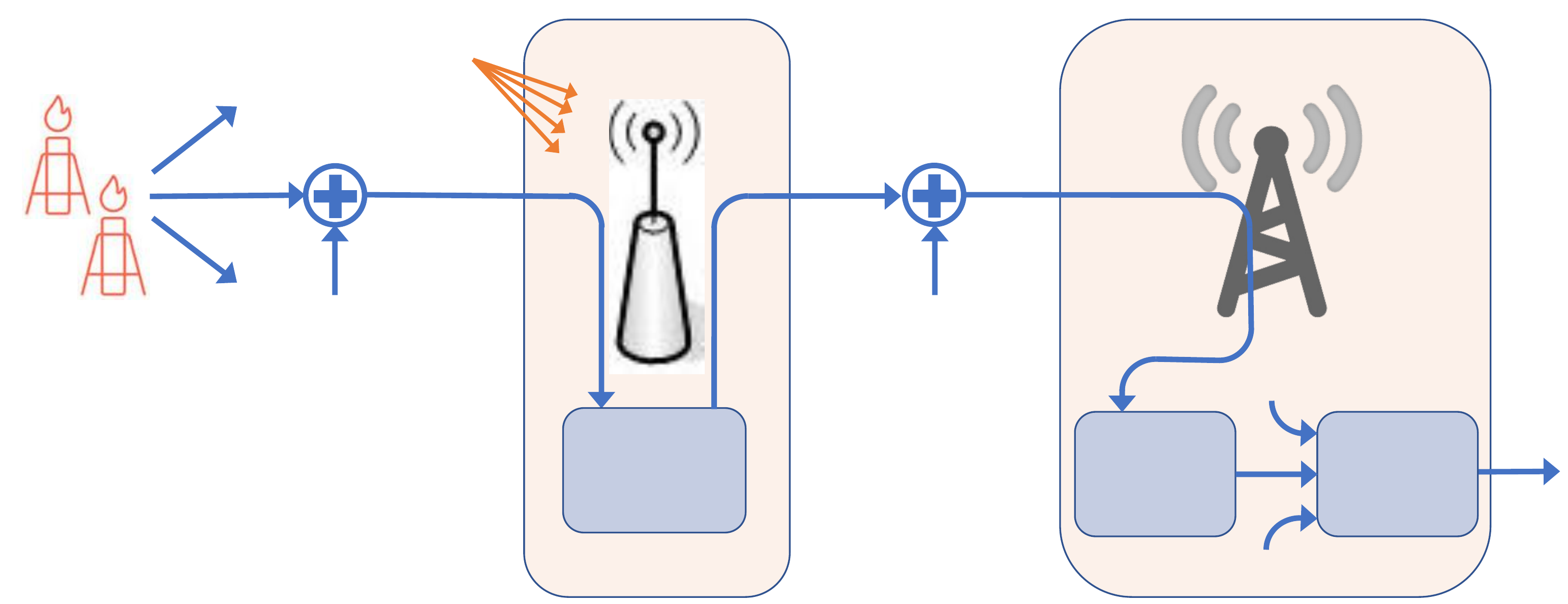}};
	\draw (-6.9, -.3) node { Unknown };
	\draw (-6.9, -.65) node {source(s) };
	\draw (-6.9, -1) node {parameters (${\boldsymbol{ \theta}}$)};
	\draw (-5.6, 2.3) node {${\bf h}_1^T {\boldsymbol{ \theta}}$};
	\draw (-5.7, 1.3) node {${\bf h}_l^T {\boldsymbol{ \theta}}$};
	\draw (-5.6, -.1) node {${\bf h}_L^T {\boldsymbol{ \theta}}$};
	\draw (-4.55, -.15) node {$v_l $};
	\draw (-3.5, 1.3) node {$x_l $};
	
	\draw (-3.5, 2.4) node {{\color{RoyalBlue}EH}};
	\draw (-1.35, 2.5) node {{\color{RoyalBlue} {\bf IoT device}}};
	\draw (-1.3, -1.5) node {{\color{RoyalBlue} {\bf Quantize}}};	
	\draw (-1.3, -2) node {{\color{RoyalBlue} {\bf \& encode}}};
	
	\draw (.5, 1.3) node {${\bf \tilde{x}}_{l,k,b} $};
	\draw (1.55, -.15) node {${\boldsymbol{\phi}}_l $};
	\draw (2.3, 1.3) node {${\bf \tilde{y}}_{l,k,b} $};
	
	\draw (4.9, 2.5) node {{\color{RoyalBlue} {\bf Fusion Center}}};
	\draw (3.8, -1.7) node {{\color{RoyalBlue} {\bf  Decode}}};
	\draw (6.3, -1.7) node {{\color{RoyalBlue} {\bf  Estimate}}};
	\draw (5.0, -.9) node {$y_{1,0,1}$};
	\draw (5.06, -1.57) node {$y_{l,k,b}$};
	\draw (5.0, -2.7) node {$y_{L,K,B}$};
	\draw (7.5, -1.5) node {$\hat{\boldsymbol \theta}$};
	\end{tikzpicture}
	\caption{System Model. The IoT device includes the sensor, the EH equipment and the wireless transmission system at ${\bf p}_l$.}\label{fig:sys_mdl_B}
\end{figure*}
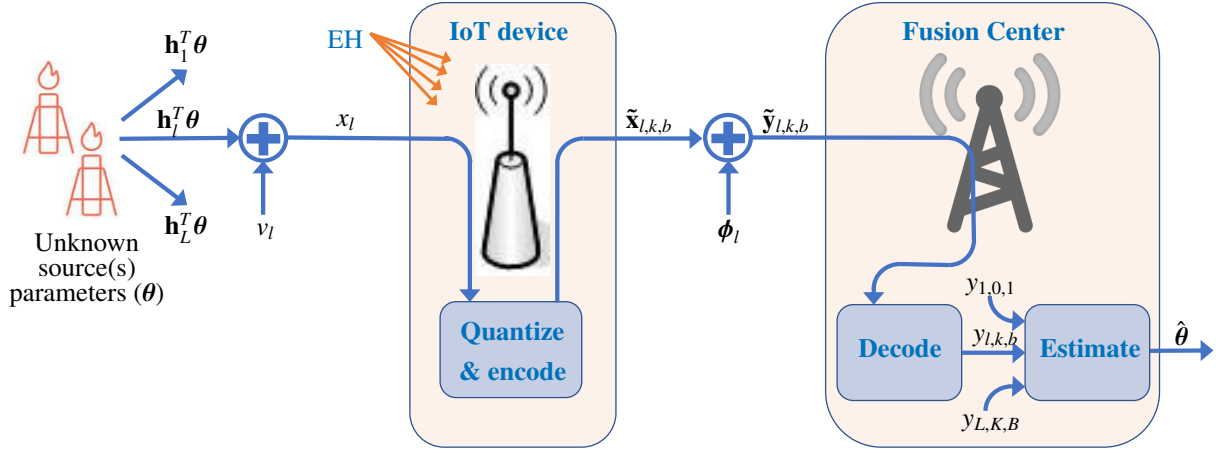

Instead of sending observations directly, in practice, each sensor in the digital transmission scheme quantizes, encodes and then transmits its observations. An illustration of the system model is shown in Figure \ref{fig:sys_mdl_B}. The measured observation, $x_l$, at $ \mathcal{S}_{(l,k,b)}$ is quantized to $2^{{N}_b R_{l,k,b}}$ levels during each transmission interval, $T$, where $R_{l,k,b}$ is the transmission rate per channel which is bounded by the Shannon capacity theorem as
\begin{equation}\label{eq4}
R_{l,k,b} \leq C_{l,k,b} = \log_{2}(1+{\rm SNR}_{l,k,b}),
\end{equation}
where, $C_{l,k,b}$ is the channel capacity. Note that $R_{l,k,b} $ and ${N}_b R_{l,k,b}$ are not necessarily integers. However, the number of quantization levels, $2^{{N}_b R_{l,k,b}}$, must be an integer.

\underline{Example:} A signal might be quantized to 9 levels and sent over one 9-ary channel or two 3-ary channels. In the first case, ${N}_b=1$ and $R_{l,k,b}= \log_2(9)$ while in the other case, ${N}_b=2$ and $R_{l,k,b} = \log_2(3)$. 

For Gaussian sensor observations, quantization distortion is given by the rate distortion theorem as \cite{cover2006elements},
\begin{equation}\label{eq2}
\sigma_{q(l,k,b)}^2 = \sigma_{x(l)}^2 2^{-2{N}_b R_{l,k,b}} 
\end{equation}
where $\sigma_{x(l)}^2$ is as defined in \eqref{eq:var_x}. To minimize distortion, we select the maximum number of quantization levels to represent each observation while $R_{l,k,b} \leq C_{l,k,b}$, i.e., the number of quantization levels is
\begin{equation}\label{eq5.5}
Q = 2^{{N}_b R_{l,k,b}} = \floor{2^{{N}_b \log_2(1+{\rm SNR}_{l,k,b})}},
\end{equation}  
where $\floor{\cdot}$ is the floor function. From \eqref{eq:SNR}, \eqref{eq2} and \eqref{eq5.5} the quantization distortion is expressed as,
\begin{align}
\sigma_{q(l,k,b)}^2 &=  \sigma_{x(l)}^{2}2^{-2\log_2\floor{2^{{N}_b \log_2(1+ {\rm SNR}_{l,k,b})}}} \nonumber \\
&= \sigma_{x(l)}^{2} \floor{(1+{\rm SNR}_{l,k,b})^{{N}_b}}^{-2} \nonumber \\
\label{eq:quantization_noise_fixed_bw}
&= \sigma_{x(l)}^2  \left\lfloor{\left(1+  \frac{P_{l,k} g_l  }{ \kappa \Delta w_b} \right)}^{{N}_b}\right\rfloor^{-2}.
\end{align}

{\bf Remark:} Unlike the analog scheme, in which by increasing the transmission bandwidth the estimation performance is not improved, the quantization distortion is decreased as the selected bandwidth $w_b$ is increased. As $w_b \to \infty$ we reach the minimum quantization error given by, 
\begin{equation}\label{eq19}
\sigma_{q(l,k)}^2 = \sigma_{x(l)}^{2} \left\lfloor \exp \left( \frac{P_{l,k}g_l N}{\kappa \Delta W} \right) \right\rfloor^{-2} ,
\end{equation}
which is obtained by applying the identity, $ \exp(a) = \lim_{b\to\infty} (1+ \frac{a}{b})^{b}$ on \eqref{eq:quantization_noise_fixed_bw}.

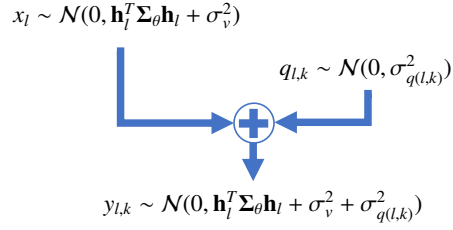
\begin{figure}
	\centering
\begin{tikzpicture}[thick,scale=0.9, every node/.style={transform shape}]
\draw (0, 0) node[inner sep=0] {\includegraphics[width=7.5cm]{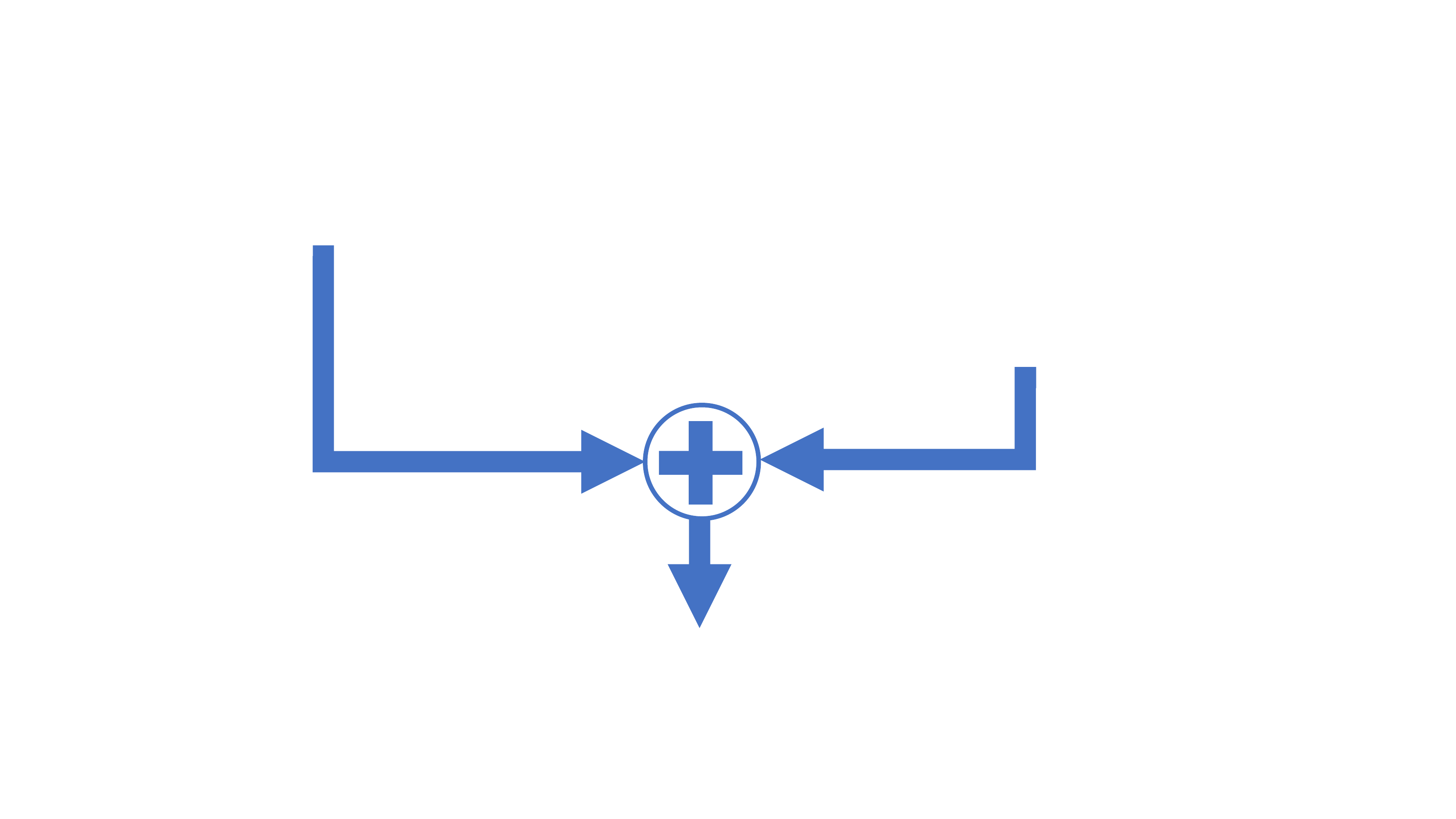}};
\draw (-2, 1.3) node {$x_l \sim \mathcal{N}(0,{\bf h}_l^T \boldsymbol{ \Sigma}_\theta {\bf h}_l + \sigma_{v}^2)$};
\draw (1.5, .5) node {$q_{l,k} \sim \mathcal{N}(0,\sigma_{q(l,k)}^2)$};
\draw (0, -1.5) node {$y_{l,k} \sim \mathcal{N}(0,{\bf h}_l^T \boldsymbol{ \Sigma}_\theta {\bf h}_l + \sigma_{v}^2 + \sigma_{q(l,k)}^2)$};
\end{tikzpicture}
\caption{Quantization error \cite{cover2006elements}.}\label{fig:quantization_error}
\end{figure}

The quantization distortion can be represented by a zero mean Gaussian signal, denoted as $q_{l,k,b}$, with variance $\sigma_{q(l,k,b)}^2$ added to the quantized signal  \cite{cover2006elements}. Figure \ref{fig:quantization_error} illustrates the quantization error.

After quantization, the observation is encoded to be sent over ${N}_b$ channels with an average power $\hat{P}_{l,k,b}$. Denoting the encoded signal as, ${\bf \tilde{x}}_{l,k,b} = [\tilde{x}_{l,k,b}^{(1)} \tilde{x}_{l,k,b}^{(2)} \dots \tilde{x}_{l,k,b}^{({N}_b)}]^T$, and assuming AWGN channels between $\mathcal{S}_{l,k,b}$ and the FC, the received signal at the FC is formulated as,
\begin{equation}\label{eq7}
\tilde{\bf y}_{l,k,b} = s_{l,k,b} (\sqrt{g_l}\tilde{\bf x}_{l,k,b}+{\boldsymbol \phi}),
\end{equation}
where, ${\bf \tilde{y}}_{l,k,b} = [\tilde{y}_{l,k,b}^{(1)} \tilde{y}_{l,k,b}^{(2)} \dots \tilde{y}_{l,k,b}^{(n_b)}]^T$ is the vector of received signals from the ${N}_b$ channels between $\mathcal{S}_{l,k,b}$ and the FC. It is assumed that ${N}_b$ is large enough such that it is possible to use  efficient coding and modulation techniques to decode the received signals with negligible error at the FC.

Considering both the measurement distortion and quantization distortion, the decoded signal received from $\mathcal{S}_{l,k,b}$ at the FC can be written as,
\begin{align}\label{eq:received_digital}
y_{l,k,b} &=  s_{l,k,b}({\bf h}_l^T {\boldsymbol \theta}+v_l+ q_{l,k,b}) \nonumber \\
&=  s_{l,k,b}({\bf h}_l^T {\boldsymbol \theta}+\tilde{e}_{l,k,b}).
\end{align}
Since $v_l$ and $q_{l,k,b}$ are two Gaussian random variables, $\tilde{e}_{l,k,b}$ is also a Gaussian random variable with zero mean and variance,
\begin{equation}\label{eq:equivalent_noise_fixed_bw}
\sigma_{\tilde{e}(l,k,b)}^2 = \sigma_{v}^2 + \sigma_{q(l,k,b)}^2. 
\end{equation}

At the FC, all selected sensors' observations are collected to estimate the unknown parameter, ${\boldsymbol\theta}$. Since the received signal at the FC can be expressed as a linear function of the unknown parameter as in \eqref{eq:received_digital}, the MMSE error covariance matrix is expressed as, \cite{kay1993fundamentals},
\begin{equation}
\boldsymbol{ \Sigma}_{{\boldsymbol\theta}|{\bf y}}({\bf S}) =\left( {\boldsymbol \Sigma}_\theta^{-1}+\sum\limits_{l =1}^{L} \sum\limits_{k =0}^{K} \sum\limits_{b =1}^{B} \dfrac{ s_{l,k,b}}{ \sigma_{\tilde{e}(l,k,b)}^2    }   {\bf h}_l {\bf h}_l^T   \right)^{-1}. \nonumber
\end{equation}	
where $\sigma_{\tilde{e}(l,k,b)}^2$ is as in \eqref{eq:equivalent_noise_fixed_bw}. Similar to the previous section, we express the estimation error by taking the trace of the error covariance matrix, 
\begin{equation}
tr \{	\boldsymbol{\Sigma}_{{\boldsymbol\theta}|{\bf y}}({\bf S}) \}=  tr\left( {\boldsymbol\Sigma_{\theta}}^{-1} + \sum\limits_{l =1}^{L} \sum\limits_{k =0}^{K}  \sum\limits_{b =1}^{B} \dfrac{ s_{l,k,b}}{ \sigma_{\tilde{e}(l,k,b)}^2 }   {\bf h_l} {\bf h_l}^T \right)^{-1}.
\end{equation}

Having the mathematical expression for the MMSE estimation error in terms of different system parameters, we are ready to define and solve \textit{the sensor "Bandwidth, Location and Power Selection for Static source estimation " (Static Source BLoPS)} problem for the digital transmission scheme. In the \textit{Static Source BLoPS} problem, the error is minimized subject to constraints on the total system cost and the total system bandwidth. The bandwidth allocation is done optimally such that each selected sensor occupies a bandwidth from the vector ${\bf w} = [w_1 \cdots w_B]^T$.

Given $L$ candidate sensor locations and $K$ sensor types with different energy harvesting capabilities, battery capacities and costs, and $B$ operating bandwidths, we would like to jointly choose the optimal subset of $\mathcal{S} = \{\mathcal{S}_{(l,k,b)} \, | \, \forall \{l,k,b\}\}$ that minimizes $tr \{	\boldsymbol{\Sigma}_{{\boldsymbol\theta}|{\bf y}}({\bf S}) \}$ subject to thresholds on the system cost and bandwidth. The \textit{Static Source BLoPS} optimization problem is mathematically written as,

{\bf Problem 2: \textit{ Static Source BLoPS:}}
\begin{align}
\underset{\{s_{l,k,b}\}}{\arg \min}
& \quad  tr \{	\boldsymbol{\Sigma}_{{\boldsymbol\theta}|{\bf y}}({\bf S}) \}
\label{eq16}\\
\text{subject to}
& \quad  \sum \limits_{l=1}^{L}{   {\bf c}^T \bf S}_l {\bf 1}_{B} \leq \lambda
\tag{\ref{eq16}a} \label{eq16a}\\
& \quad  \sum \limits_{l=1}^{L} [0 \; {\bf 1}_{K}^T ]  {\bf S}_l {\bf w} \leq W
\tag{\ref{eq16}b} \label{eq16b}\\
& \quad s_{l,k,b}  \in \{0,1\}, \quad {\forall l, k, b}
\tag{\ref{eq16}c} \label{eq16c}\\
& \quad || {\bf S}_l ||_0 =1, \quad {\forall l},
\tag{\ref{eq16}d} \label{eq16d}
\end{align}
where $|| {\bf S}_l ||_0$ is defined as the number of non-zero elements in ${\bf S}_l $.

The objective function is convex w.r.t. ${\bf S}$ \cite{chepuri2015sparsity}. The constraints \eqref{eq16a} and \eqref{eq16b} bound the system cost and bandwidth to $\lambda$ and $W$ respectively. The constraints \eqref{eq16c} and \eqref{eq16d} guarantee that the selection indicators are either zero or one and that at most one sensor is deployed at each sensing location. 
For example, the element at the $(k+1)$-th row and the $b$-th column is set to one while all other elements are equal to zero if the sensor $\mathcal{S}_{l,k,b}$ is selected. In case no sensor is selected, any element in the first row is set to one while all other elements are equal to zero. Note how the first row of ${\bf S}_l$ is excluded from the bandwidth constraint \eqref{eq16b}.

The constraints \eqref{eq16c} and \eqref{eq16d} are not convex. To obtain a convex problem, the nonconvex Boolean constraints in \eqref{eq16c} and the $\ell_0$ norm in \eqref{eq16d} are relaxed as follows,
\begin{align}
\underset{\{s_{l,k,b}\}}{\arg \min}
& \quad tr \{	\boldsymbol{\Sigma}_{{\boldsymbol\theta}|{\bf y}}({\bf S}) \}
\label{eq17}\\
 \text{subject to}
& \quad \text{constraints } \eqref{eq16a} \text{ and } \eqref{eq16b}
\tag{\ref{eq17}a} \label{eq17a}\\
& \quad s_{l,k,b}  \in [0,1], \quad {\forall l, k, b}
\tag{\ref{eq17}b} \label{eq17b}\\
& \quad ||  {\bf S}_l||_1 =1, \quad {\forall l},
\tag{\ref{eq17}c} \label{eq17c}
\end{align}
where $||{\bf S}_l||_1$ is defined as the summation of the modulus of all entries of ${\bf S}_l$. 

The optimization problem \eqref{eq17} is solved using well-known convex optimization tools. However, the solution is in general not Boolean. Therefore, a rounding algorithm should be applied to approximate the solution.

\section{Dynamic Source}
\label{dynamic}
In nature, physical quantities tend to change slowly over time. Therefore, exploiting the temporal correlation between measurements significantly improves the estimation quality. The Kalman filter, \cite{citeulike:347166}, is used at the FC to obtain the MMSE parameter estimation based on the received observations from the selected sensors over time. As opposed to the previous section which considered a static \textit{vector} source, we now focus on a dynamic \textit{scalar} source.\footnote{ We only consider the dynamic \textit{scalar} source due to the difficulty of dealing with the discrete algebraic Riccati equation (DARE) which arises from solving for the MMSE Kalman estimation error covariance matrix.}

The dynamics of the unknown parameter are captured through the first order Gauss-Markov process, i.e.,
\begin{align} 
x_l[t] &= { h}_l { \theta}[t]+v_l[t], \\
{{ \theta}}[t] &= {a} {{ \theta}}[t-1] +{u}[t], \quad t \in \mathbb{Z}_{++}
\end{align}
where $\mathop{\mathbb{E}} \{\theta[0]\} = \mu_s $ and $u[t] \sim \mathcal{N}(0,\sigma_{u}^2)$ is the driving or excitation noise. We assume that $\theta[0]$ and $u[t]$ are independent and, $u[t_1]$ and $ u[t_2]$ are uncorrelated $\forall t_1 \neq t_2$. For a stabilizable  $(a,\sigma_{u})$ and as $t \to \infty$,  $\mu_{\theta} = \lim\limits_{t \to \infty} \mathop{\mathbb{E}} \{\theta[t]\} = 0 $ and $$\sigma_{\theta}^2 =\lim\limits_{t \to \infty} \mathrm{Var}(\theta[t]) = \sigma_u^2/(1-a^2).$$ Since the selection is done at the design time, we consider the steady state Kalman MMSE estimation. In the following subsections, we study the sensor selection for the analog and digital transmission schemes.

\subsection{Analog Transmission Scheme}

Following the same analog scheme derivations as in the previous section, the received observation from $\mathcal{S}_{l,k,b}$ at the FC is expressed as
\begin{align*}
y_{l,k,b} = s_{l,k,b} \bigg(\frac{\sqrt{\hat{P}_{l,k,b}g_l} x_l}{\sigma_{x(l)} }  + \phi_l \bigg),
\end{align*}
where $x_l = h_l \theta + v_l $ is the scalar measurement observation at the sensor and $\sigma_{x(l)}^2 = \mathbb{E}\{x_l^2\} $. Note that we drop the time index for simple presentation. Without loss of generality, the received signal is normalized as
\begin{eqnarray}
y_{l,k,b} &=& s_{l,k,b} ({ h}_l { \theta} + e_{l,k,b} ),
\end{eqnarray} 	
where, $e_{l,k,b} = v_l +\dfrac{\phi_l \sigma_{x(l)}}{\sqrt{\hat{P}_{l,k,b} g_l}} $ is the equivalent noise. $e_{l,k,b}$ is a zero mean Gaussian variable with variance,

\begin{eqnarray}
\sigma_{e(l,k,b)}^2 = \sigma_v^2 + \dfrac{({ h}_l {\Sigma_{\theta}} { h}_l + \sigma_v^2 )\sigma_{\phi}^2}{g_l \hat{P}_{l,k,b}}.
\end{eqnarray}	
 Define ${\boldsymbol \epsilon}({\bf S})= [\epsilon_1 \cdots \epsilon_L ]^T$ with $ \epsilon_l = \sum_{k=0}^{K }\sum_{b=1}^{B} {s}_{l,k,b} \, e_{l,k,b} $ where ${\bf S}$ is the selection indicator set of matrices, ${\bf S}_l, \; \forall l$, as defined before. Now, the received vector of observations from all selected sensors at the FC is formulated as,
 
\begin{equation}\label{eq:received_digital_vector}
{\bf y}({\bf S}) = {\bf h} \theta + {\boldsymbol \epsilon}({\bf S}).
\end{equation}
Here, ${\bf y}({\bf S}) \in \mathbb{R}^L$ represents the received observations from all the sensing locations as a function of the sensor type and bandwidth selection at each location. ${\bf h} = [h_1 \cdots h_L]^T$ represents the vector of measurement gains over sensing locations. Since, $e_{i,k,b}$ and $e_{j,k,b}$ are uncorrelated for any $i \neq j$, the elements of ${\boldsymbol \epsilon}({\bf S})$ are uncorrelated. Consequently, the covariance matrix of ${\boldsymbol \epsilon}({\bf S})$, denoted as ${\boldsymbol{ \Sigma}}_{\epsilon}({\bf S})$ is diagonal such that, \\
\begin{align} \label{eq:covariance_epsilon}
[{\boldsymbol{ \Sigma}}_{\epsilon}({\bf S})]_{ll} = \mathbb{E} \{ \epsilon_l^2 \} =\sum_{k=0}^{K }\sum_{b=1}^{B} {s}_{l,k,b} \, \sigma_{e(l,k,b)}^2 . 
\end{align}
Since only one sensor is selected at any location, only one term of the summation is non zero. To avoid the indefinite form $e_{l,0,b} s_{l,0,b}= \infty \cdot 0 $ that arises with the auxiliary sensor type (with $\hat{P}_{l,0,b}=0$ ) not being selected, we redefine the $t_0$ transmission power as $\hat{P}_{l,0,b}\approx 0$.

Assuming that $(a,\sigma_{u})$ is stabilizable, the MMSE Kalman estimation error as $t \to \infty$ converges to \cite{kay1993fundamentals,anderson1979optimal}
\begin{align}
M({\bf S}) &= M({\bf S})[t]  \\
&= \Big[ 1- {\bf h}^T  ( \dfrac{{\boldsymbol{ \Sigma}}_{\epsilon}({\bf S})}{M_p({\bf S})}  + {\bf h} {\bf h}^T )^{-1} {\bf h}     \Big] M_p({\bf S}), \label{eq:kalman_mmse} 
\end{align}
where $M_p({\bf S}) = M_p({\bf S})[t]= a^2 M({\bf S})[t-1] + \sigma_{u}^2 $ is the MMSE Kalman prediction error. Since the MMSE Kalman estimation error converges as $t \to \infty$, $M({\bf S})=M({\bf S})[t-1]$. Therefore, the MMSE Kalman prediction error is expressed as,
\begin{align}\label{eq:kalman_prediction}
M_p({\bf S})= a^2 M({\bf S}) + \sigma_{u}^2.
\end{align} 
The MMSE Kalman estimation error can be derived by substituting \eqref{eq:kalman_prediction} into \eqref{eq:kalman_mmse}.

For the analog transmission scheme, increasing the transmission bandwidth is unnecessary as proved in Appendix A. Therefore, the selection is reduced to one operating bandwidth, $w_b = \frac{W}{N}$. The MMSE Kalman estimation error is minimized subject to a prescribed system budget and bandwidth by solving the \textit{Dynamic Source LoPS} optimization problem expressed as, 

{\bf Problem 3: \textit{ Dynamic Source LoPS:}}
\begin{align}
 \underset{\{s_{l,k}\}}{\arg\min}
& \quad   M({\bf s}) 
\label{eq:non_convex}\\
 \text{subject to}
& \quad \sum \limits_{l=1}^{L} {\bf c}^T {\bf s}_l \leq \lambda 
\tag{\ref{eq:non_convex}a} \label{eq:non_convex_a}\\
& \quad  \sum \limits_{l=1}^{L} [0 \; {\bf 1}_{K}^T ] {\bf s}_l \leq N
\tag{\ref{eq:non_convex}b} \label{eq:non_convex_b}\\
& \quad s_{l,k}  \in \{0,1\}, \quad {\forall l, k}
\tag{\ref{eq:non_convex}c} \label{eq:non_convex_c}\\
& \quad || {\bf s}_l||_0 = 1, \quad {\forall l}.
\tag{\ref{eq:non_convex}d} \label{eq:non_convex_d}
\end{align}

Neither the objective function in \eqref{eq:non_convex} nor the constraints \eqref{eq:non_convex_c} and \eqref{eq:non_convex_d} are convex. Hence, the optimization problem cannot be efficiently solved using well-known methods \cite{boyd2004convex}.

{\bf Proposition 3:}
The minimization of $M({\bf S}) $ is equivalent to maximizing $\gamma({\bf S})$ where,
\begin{eqnarray}\label{eq:gamma_s}
\gamma({\bf S}) = \sum_{l=1}^{L} \sum_{k=0}^{K} \sum_{b=1}^{B}  \dfrac{(h_l)^2 g_l \hat{P}_{l,k,b}  }{\sigma_v^2  g_l \hat{P}_{l,k,b} +\sigma_{x(l)}^2\sigma_{\phi}^2} s_{l,k,b}
\end{eqnarray}

{\bf Proof:} See Appendix \ref{app:min_M_to_max_gamma}.

By replacing $M({\bf S}) $ with $\gamma({\bf S})$ and relaxing the constraints \eqref{eq:non_convex_c} and \eqref{eq:non_convex_d}, the convex (linear) optimization problem can be expressed as,
\begin{align}
 \underset{\{s_{l,k}\}}{\arg\max}
& \quad   {\gamma({\bf s}})
\label{eq:convex_3}\\
 \text{subject to}
& \quad \eqref{eq:non_convex_a} \eqref{eq:non_convex_b}
\tag{\ref{eq:convex_3}a} \label{eq:convex_3_a}\\
& \quad s_{l,k}  \in [0,1], \quad {\forall l,k}
\tag{\ref{eq:convex_3}b} \label{eq:convex_3_b}\\
& \quad {\bf 1}^T {\bf s}_l  =1, \quad {\forall l}.
\tag{\ref{eq:convex_3}c} \label{eq:convex_3_c}
\end{align}

Equivalently, we can minimize the system cost subject to a prescribed MMSE reconstruction error, $M({\bf S})$. The relaxed equivalent problem is written as,
\begin{align}
 \underset{\{s_{l,k}\}}{\arg\min}
 & \quad  \sum \limits_{l=1}^{L} {\bf c}^T {\bf s}_l
\label{eq:convex}\\
 \text{subject to}
 & \quad {\gamma({\bf S}}) \geq \tilde{\xi}
\tag{\ref{eq:convex}a} \label{eq:convex_a}\\
& \quad \text{constraints } \eqref{eq:non_convex_b}, \eqref{eq:convex_3_b} \text{ and } \eqref{eq:convex_3_c}
\tag{\ref{eq:convex}b} \label{eq:convex_b}
\end{align}
where bounding $\gamma({\bf S})$ to be greater than $\tilde{\xi}$ is equivalent to bounding $M({\bf S})$ to be less than $\xi$ such that,
\begin{eqnarray}
\tilde{\xi} = \dfrac{ \sigma_u^2 -   (1-a^2) \xi }{a^2 \xi^2 + \sigma_u^2 \xi}
\end{eqnarray}
This is proved by substituting $M({\bf S})$  and $\gamma({\bf S})$  with $\xi$ and $\tilde{\xi}$ in \eqref{eq:app_d} respectively, and since $M({\bf S})$ and $\gamma({\bf S})$ are inversely proportional.

\subsection{Digital Transmission Scheme}

In the digital transmission scheme, the sensors' observations are quantized and encoded such that the transmission to the FC is error free. Following the same derivations as in the previous section, the decoded vector of received signals at the FC is as expressed in \eqref{eq:received_digital_vector} with the ${\boldsymbol{\epsilon}({\bf S} ) } $ covariance matrix redefined as,

\begin{align}\nonumber
& [{\boldsymbol{ \Sigma}}_{\epsilon}({\bf S})]_{ll} = \\ \nonumber
& \sum_{k=0}^{K} \sum_{b=1}^{B} \big( \sigma_v^2 + \sigma_{x(l)}^2  \left\lfloor{\left(1+ \frac{P_{l,k} g_l  }{ \kappa \Delta w_b} \right)}^{{N}_b}\right\rfloor^{-2} \big) s_{l,k,b}.
\end{align}

Following a similar derivations to the analog case, it can be shown that minimizing the steady state Kalman MMSE estimation error, $M({\bf S}) $, is equivalent to maximizing 
\begin{align}\nonumber
\gamma({\bf S}) &=  {\bf h}^T  {\boldsymbol{ \Sigma}}_{\epsilon}({\bf S})^{-1}   {\bf h} = \sum_{l=1}^L \dfrac{ {h}_l^2}{[{\boldsymbol{ \Sigma}}_{\epsilon}({\bf S})]_{l,l}} .
\end{align}
Since $s_{l,k,b} \in \{0,1\}$ and $ || {\bf S}_l ||_0 = 1$, $\gamma({\bf S})$ can be written as,

\begin{align}
\gamma({\bf S}) = \sum_{l=1}^L \sum_{k=0}^{K} \sum_{b=1}^B \dfrac{h_l^2 s_{l,k,b} }{  \Bigg( \sigma_v^2 + \sigma_{x(l)}^2  \left\lfloor{\left(1+ \frac{P_{l,k} g_l  }{ \kappa \Delta w_b} \right)}^{{N}_b}\right\rfloor^{-2} \Bigg) }.
\end{align}

Now, having a convex formulation for the Kalman MMSE error to be minimized (through an equivalent maximization of a linear function), we are ready to formulate the \textit{dynamic source BLoPS} optimization problems.

The relaxed optimization problem for selecting the sensor bandwidth, location and power to minimize the Kalman MMSE error subject to a prescribed system cost and bandwidth is formulated as,

{\bf Problem 4: \textit{ Dynamic Source BLoPS:}}
\begin{align}
 \underset{\{s_{l,k,b}\}}{\arg \max}  & \quad  { \gamma}({\bf S}) \label{eq1}\\
 \text{subject to}
& \quad \sum \limits_{l=1}^{L}{   {\bf c}^T  \bf S}_l {\bf 1}_{B} \leq \lambda
\tag{\ref{eq1}a} \label{eq1a}\\
& \quad \sum \limits_{l=1}^{L}  [0 \; {\bf 1}_{K}^T ]  {\bf S}_l {\bf w} \leq W
\tag{\ref{eq1}b} \label{eq1b}\\
& \quad s_{l,k,b}  \in [0,1], \quad {\forall l, k, b}
\tag{\ref{eq1}c} \label{eq1c}\\
& \quad ||  {\bf S}_l||_1 =1, \quad {\forall l}.
\tag{\ref{eq1}d} \label{eq1d}
\end{align}

\section{Rounding Algorithms}
\label{rounding_algorithms}

In the ideal case, the solution of the relaxed problem is $s_{l,k,b} \in \{ 0,1 \}$ with $ || {\bf S}_l ||_0 = 1$, $\forall l,k,b$. However, due to the box ($\ell_1$) relaxation, the solution is in general  $s_{l,k,b} \in [ 0,1 ]$. Rounding algorithms are needed to approximate the unfeasible solutions obtained in the previous two sections to a feasible solution with a selection indicator $s_{l,k,b} \in\{0,1\}$. Due to the constraint $||{\bf S}_l||_0=1$, the simple rounding algorithm, via the function $round(s_{l,k})$ and the randomized algorithm proposed in \cite{chepuri2015sparsity} are unreliable and lead to unfeasible solutions. 

In the conventional randomized algorithm \cite{chepuri2015sparsity}, $J$ realizations are generated where in each realization the sensor $\mathcal{S}_{l,k,b}$ is selected with probability $s_{l,k,b}$. By performing exhaustive search for the minimum MSE over the $J$ realizations a solution is obtained. The randomized algorithm outperforms the simple rounding algorithm at the expense of a higher computational cost while solving for the reconstruction MSE of $J$ realizations. The randomized algorithm proposed in \cite{chepuri2015sparsity} is suitable for the setup in which the goal is only to select sensing locations. Due to the possibility of violating the constraint $||{\bf S}_l||_0=1$, most of the realizations will be eliminated. 

To round the obtained solutions efficiently, we propose a novel rounding algorithm which takes the sensor type and bandwidth selection into consideration. The proposed randomized rounding algorithm summarized in Algorithm \ref{Randomized_algorithm}, generalizes the randomized rounding algorithm proposed in \cite{chepuri2015sparsity} by taking the extra constraints into account.

\begin{algorithm}[H]
	\caption{Randomized Rounding}
	\begin{algorithmic}[1] \label{Randomized_algorithm}	
		\STATE Let $ \Psi_l= (\Psi_{l,1},\Psi_{l,2})$, $\forall l$, be a random vector of population $ \{(0,1),\cdots, (K,B)\}$ and $\mathbb{P}\{\Psi_l = (k,b) \} = s_{l,k,b}, \; \forall l$ 
		\STATE Generate $ j= \{1, \cdots, J \}	$ realizations 
		\STATE Let $\psi_l^{(j)} = (\psi_{l,1}^{(j)},\psi_{l,2}^{(j)})$ be the $j$-th realization of $\Psi_l, \; \forall l$
		\STATE Let $\Omega = \{ j| \sum_{l=1}^L c_{\psi_{l,1}^{(j)}} \leq \lambda,  \, \sum_{l=1}^L w_{\psi_{l,2}^{(j)}} \leq W, \, \} $ be the set of all realizations that satisfy the constraints. 
		\STATE If $\Omega$ is empty, go back to step (2). 
		\STATE Define $\hat{s}_{l,k,b}^{(j)} $ such that $\hat{s}_{l,k,b}^{(j)}=1 $ if $\psi_l^{(j)} =\{k,b\}$ otherwise,  $\hat{s}_{l,k,b}^{(j)} =0$, $\forall l,k,b,j$.	
		\STATE The suboptimal Boolean solution is $ \tilde{\bf S} = \underset{j\in\Omega}{\arg\min} \quad \! tr \{	\boldsymbol{\Sigma}_{{\boldsymbol\theta}|{\bf y}}(\hat{\bf S}^{(j)}) \}$ for the static source and $\underset{j\in\Omega}{\arg\min} \quad \! \gamma(\hat{\bf S}^{(j)}) $ for the dynamic source.
	\end{algorithmic}
\end{algorithm}

\section{Numerical Experiments}
\label{num_experiments}
\begin{figure}
	\centering
	\includegraphics[width=9cm]{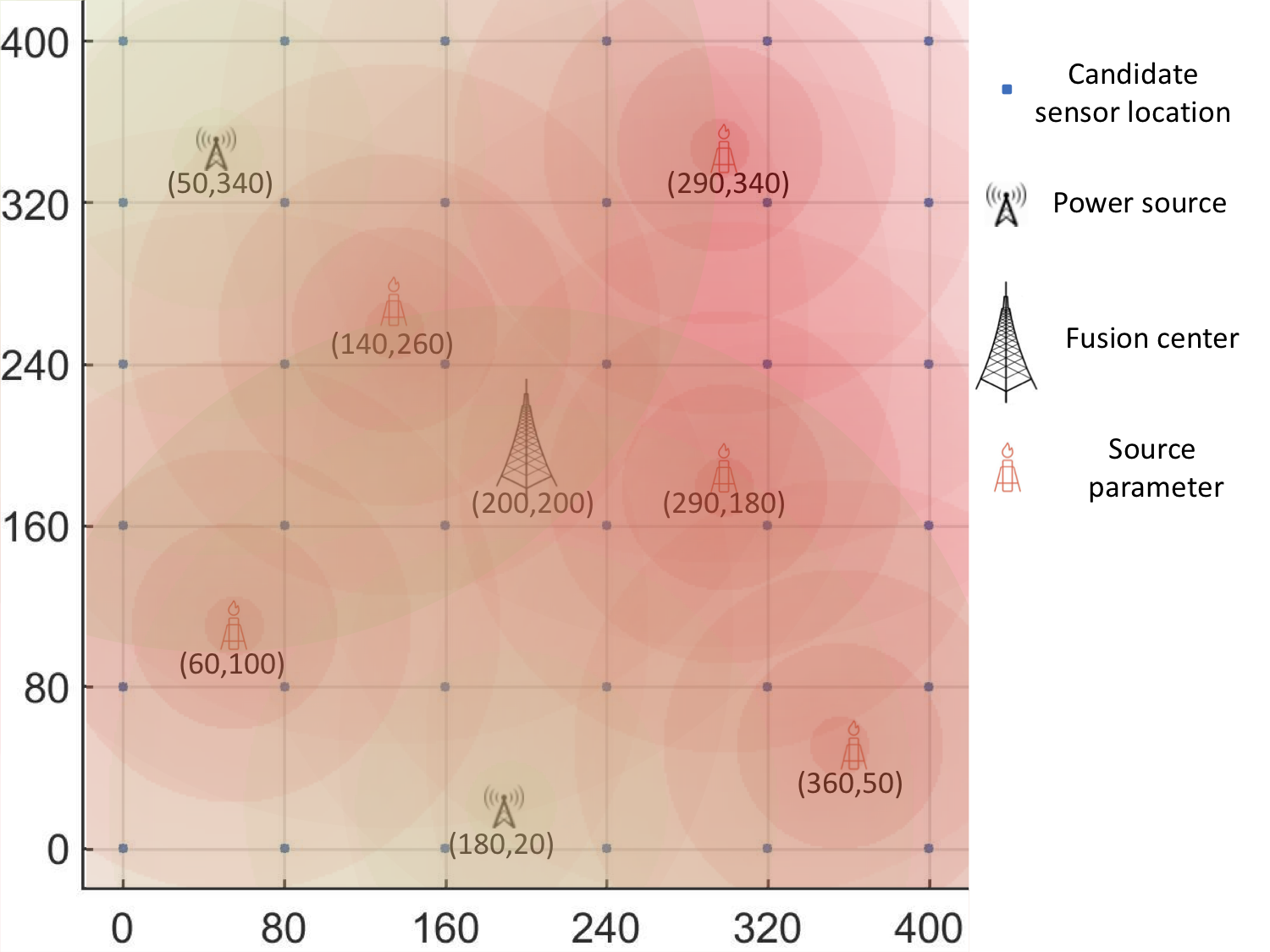}
	\caption{System map includes a grid of 36 candidate sensor locations, one FC, five sources and two BSs at different locations.}\label{fig:locations}
\end{figure}

Consider a field of area $400\times400 \,$ [m$^2$] with $L =36$ candidate sensing locations distributed uniformly. We select sensors from a pool of $K=3$ sensor types and $B=3$ operating bandwidths to be placed at a subset of the $L$ sensing locations such that $\lambda \leq 35 \,$ [k\$]  and $W \leq 1 \, $[MHz]. A measurement is collected every $T=1 \,$ [ms] at each selected sensor. The observed measurement at $\mathcal{S}_{l,k,b}$ is a linear combination of the diffused unknown parameter from $m$ sources. As a practical example for the sensor selection for source estimation, consider a chemical plant at which sensors are placed to estimate the gas emission from gas flares. Assume that the gas diffuses from the flares to the sensing locations as follows,
\begin{align}
h_{l,m} = \beta_1 \exp(-d_{l,m}/\beta_2)  \mathbbm{1}_{\{ d_{l,m} \leq \beta_3 \}},
\end{align} 
where $d_{l,m}$ is the distance between the sensing location ${\bf p}_l$ and the $m$-th source, $\beta_1, \beta_2$ are the source diffusion parameters and $\mathbbm{1}_{\{ d_{l,m} \leq \beta_3 \}}$ is the indicator function which equals one if $ d_{l,m} \leq \beta_3$ and zero otherwise. For the scalar dynamic source, we consider $m=1$ source located at $(290,180)$. While for the vector static sources, we consider $m=5$ sources located as shown in Figure \ref{fig:locations}.
The selected sensors harvest solar energy and electromagnetic energy from cellular base stations (BS) and use it to transmit their observations to the FC\footnote{hardware and signal processing powers are neglected}. The maximum energy harvesting at the location ${\bf p}_l$ is given by,
\begin{align}
\rho_{l} = \sum_{i=1}^I \varrho_i d_{i,l}^{-\alpha} +  \varrho_0
\end{align} 
where $\varrho_i$ is the $i$-th BS transmission power, $ \varrho_0$ is the EH from solar cells and $d_{i,l}$ is the distance between the $i$-th BS and ${\bf p}_l$. The source parameters, candidate sensing locations, BSs and FC are shown in Figure \ref{fig:locations}. Unless otherwise stated, the default system parameters are presented in Table \ref{table:parameters}.

\begin{table}[H]
	\caption{Default system parameters.}	\label{table:parameters} 
	\centering
	\begin{tabular}
		{| p{.5cm} | p{2.1cm}  |  p{.5cm} | p{1.3cm} |  p{.5cm} | p{1.3cm} |} 
		\hline
		Par. & Value & Par. & Value & Par. & Value \\ [0.5ex] 
		\hline\hline
		${\bf c}$ & $[0 \, 1 \, 2 \, 3] \, $[k\$]  &$ {\boldsymbol{ \Sigma}}_{\theta}$ & $ {\bf I}_5 $ & $\alpha$& $2$ \\ 
		\hline
		${\boldsymbol \eta}$ & $[0 \, .3 \, .6 \, .9]$ & $\sigma_{u}^2$ & $5$  & $\beta_1$ & $10$ \\
		\hline
		$ {\boldsymbol \varepsilon}$ & $[0 \, .3 \, .6 \, .9] \times10^{-3}$  & $\sigma_{v}^2$ & $ 1 $ & $ \beta_2$  & $ 100 $  \\
		\hline
		${\bf w}$ & $[20 \, 40 \, 60] \, $[kHz] & $\sigma_\phi^2$ & $ -60 \, $[dBm]  & $ \beta_3 $ & $250$  \\
		\hline
		a & $0.71$ & $\varrho_i $ & $1 $[dB]  & $ \varrho_0 $ & $-3 $[dBm]  \\
		\hline
	\end{tabular}
\end{table}

\begin{figure}
	\centering
	\includegraphics[width=9cm]{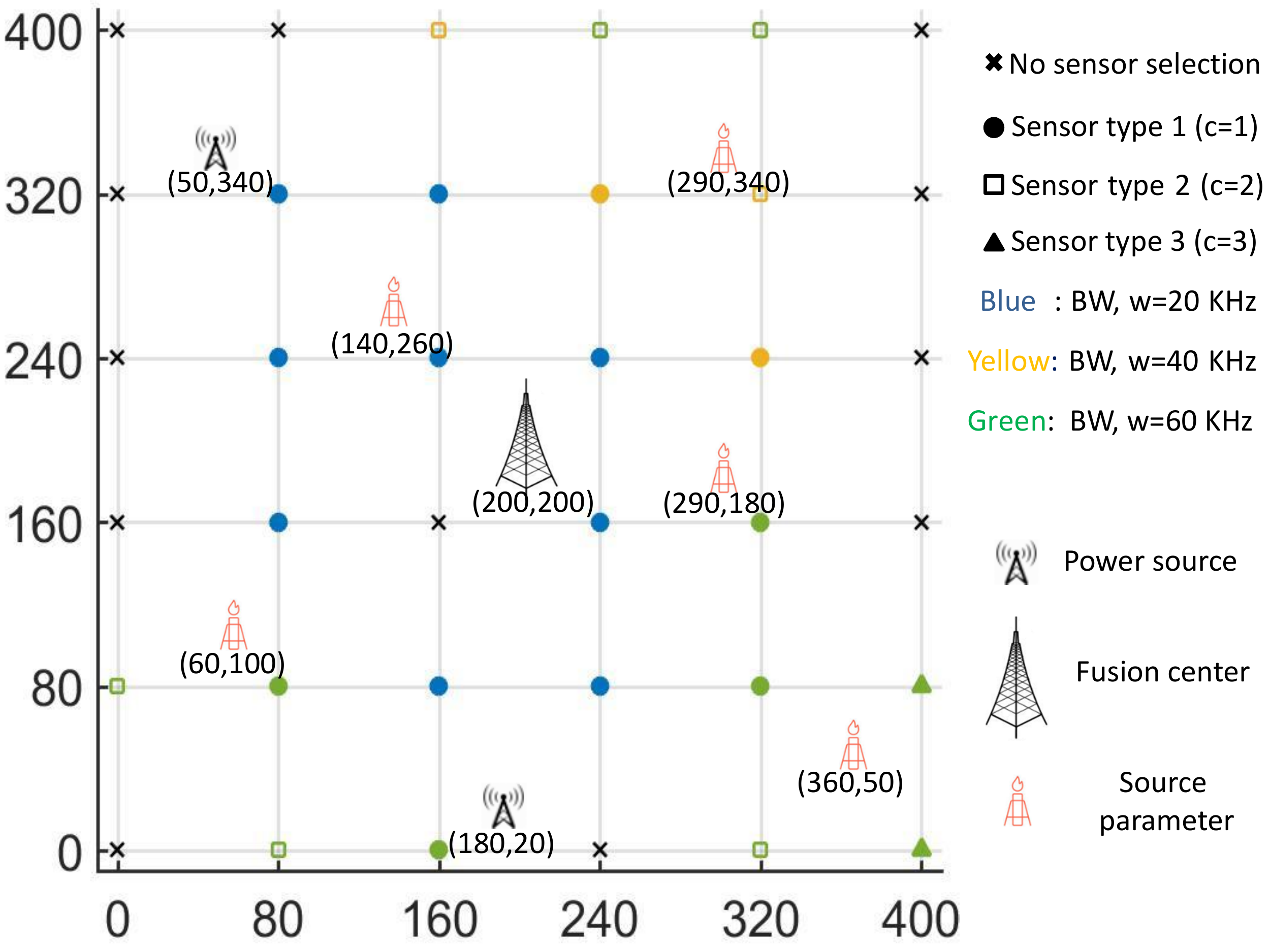}
	\caption{A sensor selection example for the Static Source BLoPS problem. ($\sigma_v^2=10^{-2.5}$).}\label{fig:selection}
\end{figure}

Using the CVX optimization tool box \cite{cvx}, we directly solve the static source LoPS and BLoPS, and the dynamic source LoPS and BLoPS problems. Then, the obtained solution is rounded using Algorithm \ref{Randomized_algorithm}. The solution of the static source BLoPS problem is shown in Figure \ref{fig:selection}. As shown in Figure \ref{fig:selection}, more expensive and higher bandwidth sensors are selected at far distances from the FC (and BSs) as compared to candidate sensor locations close to the FC (and BSs). This is expected since far sensors from the FC need higher resources to guarantee reliable communication link to the FC. Also note how most of the un-selected sensing locations are at the edge of the field.

\begin{figure}
	\centering
	\includegraphics[width=9cm]{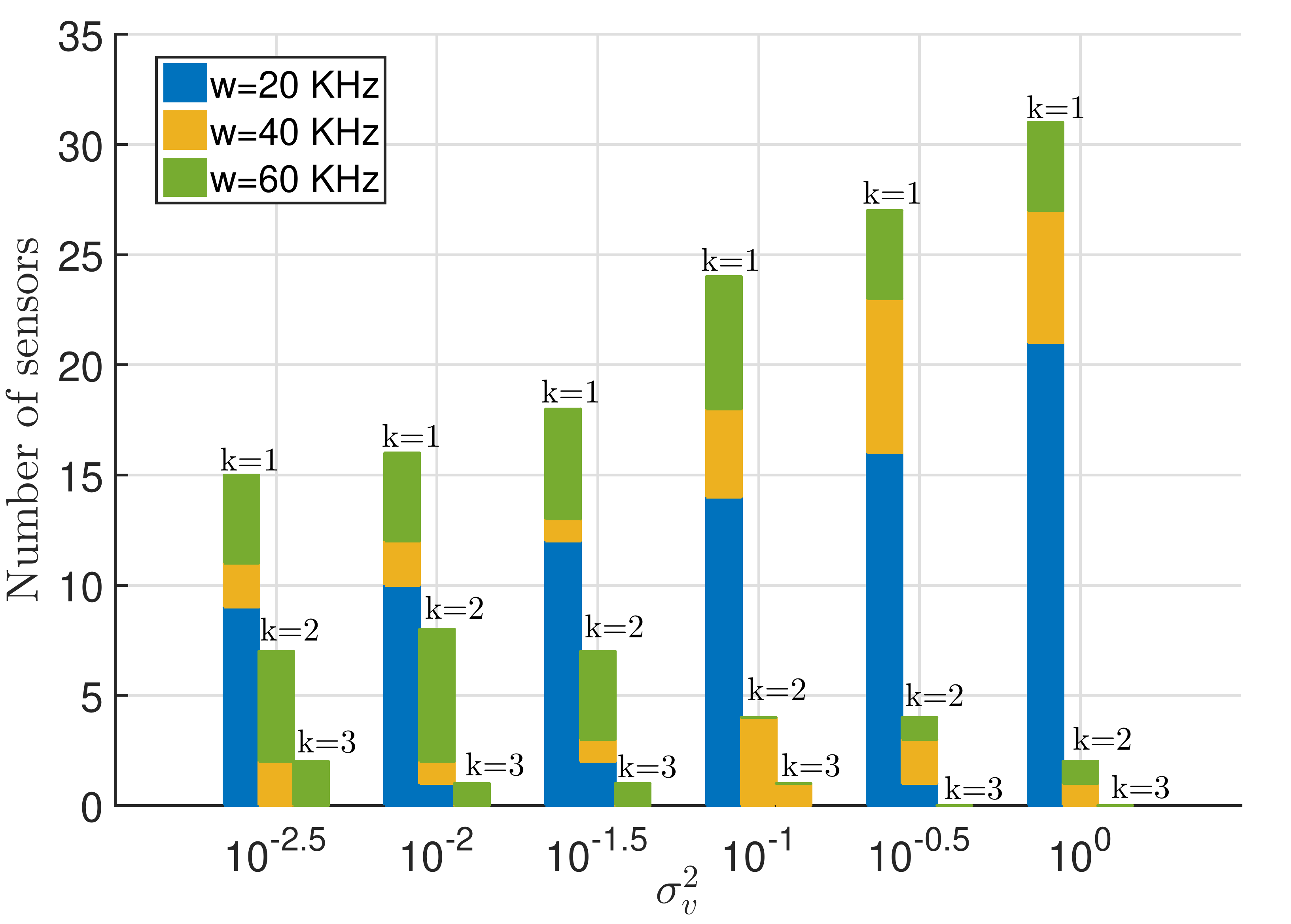}
	\caption{Number of selected sensors from each sensor type and operation bandwidth against $\sigma_v^2$.} \label{fig:number_of_sensors}
\end{figure}

The number of selected sensors from each sensor type and operating bandwidth is shown in Figure \ref{fig:number_of_sensors} against the measurements noise variance. When the measurement noise variance is low compared to the communication channel noise variance, less but more equipped sensors are selected. On the other hand, if the measurement noise is dominant over the channel noise, selecting many cheap sensors with a low operating bandwidth becomes more suitable.

\begin{figure}
	\centering
	\includegraphics[width=9cm]{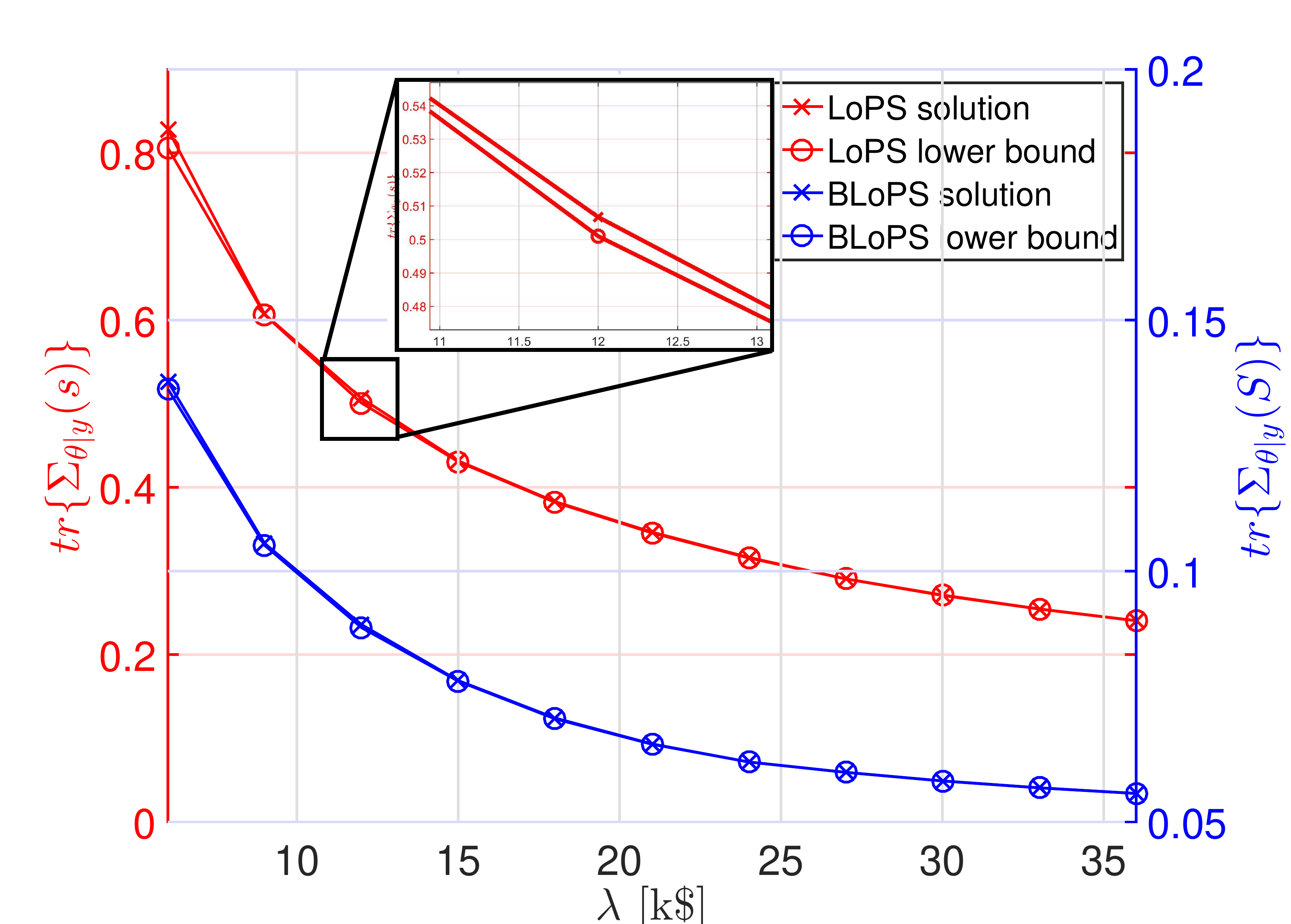}
	\caption{Reconstruction error for Static Source LoPS and BLoPS against system cost.}\label{fig:static}
\end{figure}

\begin{figure}
	\centering
	\includegraphics[width=9cm]{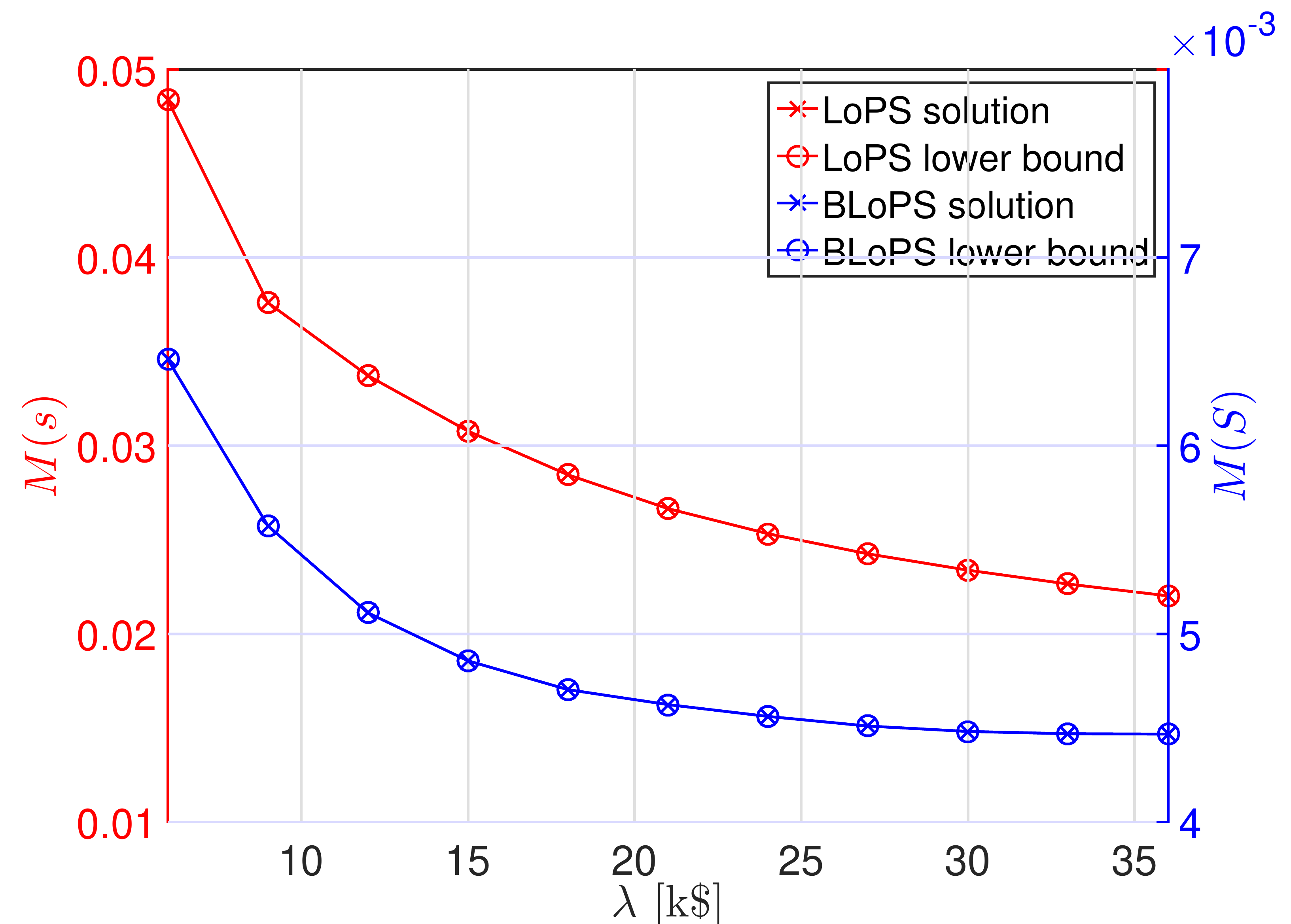}
	\caption{Reconstruction error for Dynamic Source LoPS and BLoPS against system cost.}\label{fig:dynamic}
\end{figure}

Recall that the static source LoPS and dynamic source LoPS problems are used when the analog transmission scheme is assumed while the static source BLoPS and dynamic source BLoPS problems are utilized when the digital transmission scheme is assumed. Figures \ref{fig:static} and \ref{fig:dynamic} show the obtained source estimation MMSE versus $\lambda$. The digital scheme outperforms the analog scheme at the expense of a higher system bandwidth consumption. Recall that in order to neglect the AWGN channel noise for digital transmission, ${N}_b$ must be large to enable channel coding. The solutions of the relaxed optimization problems throughout the paper represent a lower bound for the achievable estimation MMSE. As shown in Figures \ref{fig:static} and \ref{fig:dynamic}, the solution obtained using the proposed rounding algorithm is very close to the lower bound solution. This is partly because most of the selection indicators are Boolean even before rounding. 

\begin{figure}
	\centering
	\includegraphics[width=9cm]{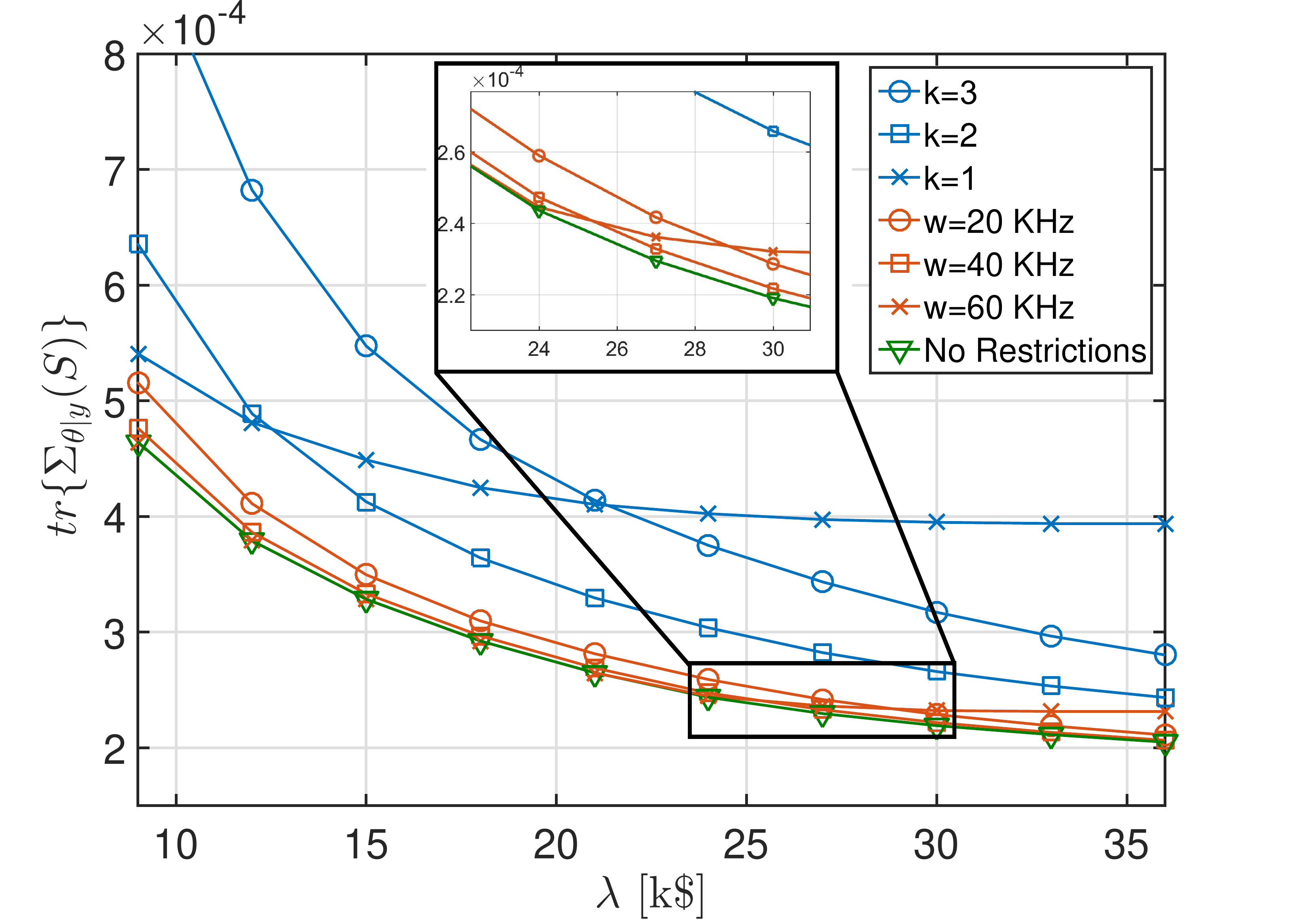}
	\caption{Reconstruction error for Static Source BLoPS with selection restrictions.} \label{fig:channelsost_kinds}
\end{figure}

Taking the static source BLoPS problem as an example, Figure \ref{fig:channelsost_kinds} shows the advantage of adding higher degrees of flexibility to the system by allowing different sensor types and operating bandwidths. In the figure, the blue curves restrict the sensor type selection while jointly optimizing sensing location and operating bandwidth. Similarly, the orange curves restrict the operating bandwidth while the sensor location and type are jointly optimized. The green curve is obtained by jointly optimizing the sensor bandwidth, location and type. From Figure \ref{fig:channelsost_kinds}, we note that restricting the system types degrades the system performance considerably while restricting the sensor transmission bandwidth is less influential. For $\lambda >12$ [K\$], restricting the sensor type to $k=2$ results in a better performance as compared to restricting the sensor type to $k=1$ and $k=3$. This is because expensive sensors might provide extra unnecessary power for sensors close to the FC while cheap sensors might be useless for sensors far from the FC (see how the curve $k=1$ decreases slowly after $\lambda = 23$ [K\$]). Finally, notice how at $\lambda \leq 18$ [K\$], the flexible solution (with all sensor types and bandwidths allowed) and the solution with all types but only bandwidth $w_b= 60$ [KHz] are identical. This is because there is enough system bandwidth to allow all selected sensors to transmit over $w_b=60$ [KHz]. As $\lambda$ increases, the number of selected sensors also increases and the smaller transmission bandwidths become more suitable for some of the selected sensors. 

\section{Conclusion}

Novel models for sensor selection have been introduced and optimized to minimize the source estimation MMSE at a central unit given a limited system cost and spectral budgets. The EH and the communication channel quality were taken into account in addition to the measurement accuracy. A digital transmission scheme between sensors and the FC was modeled based on information theory. We show that the digital transmission scheme outperforms the analog scheme given enough bandwidth to encode data efficiently. A flexible sensor selection is optimized, where not only the sensor location is selected but also the power and bandwidth. Relaxing the amount of power and bandwidth utilized to send the sensors' observations offers better inference quality at the fusion center.


\section{The effect of bandwidth increase on the analog communication}
\label{app:bandwidth_increase_analog}

Assume ${N}_b$ observation copies of a selected sensor, $\mathcal{S}_{l,k,b}$,  are transmitted over ${N}_b$ channels. The received signal from the $i$-th channel, $\forall i \in \{1, \cdots, N_b\}$,  is expressed as,
\begin{align*}
y_{l,k,b}^{(i)} = \frac{\sqrt{\hat{P}_{l,k,b}g_l} x_l}{\sigma_{x(l)} }  + \phi_l^{(i)}.
\end{align*}
The FC receiver noise is reduced by averaging the received observation copies from $\mathcal{S}_{l,k,b}$ as, 
\begin{align*}
\bar{y}_{l,k,b} =  \frac{\sqrt{\hat{P}_{l,k,b}g_l} x_l}{\sigma_{x(l)} }  +\dfrac{1}{N_b} \sum_{i=1}^{N_b} \phi_l^{(i)}
\end{align*}
Assuming independent FC receiver noises over the ${N}_b$ channels, the SNR is expressed as
\begin{align*}
{\rm SNR}_{l,k,b} &= \dfrac{ \hat{P}_{l,k,b} g_l }{\mathbb{E} \Big\{ \big(\dfrac{1}{n_b} \sum_{i=1}^{n_b} \phi_l^{(i)}\big)^2 \Big\} } \\
&= \dfrac{ \hat{P}_{l,k,b} g_l }{\sigma_{\phi}^2/n_b} =  \frac{P_{l,k} g_l  }{ \kappa \Delta } \dfrac{N}{ W}
\end{align*}
which is independent of ${N}_b$.


\section{Proof that $\min M({\bf S})$ is equivalent to $\max \gamma({\bf S}) $} 
\label{app:min_M_to_max_gamma}

From \eqref{eq:kalman_mmse} and from the matrix inversion lemma,
\begin{align}
\hat{\bf A}^{ -1} \hat{\bf B} (\hat{\bf D} -\hat{\bf C} \hat{\bf A}^{ -1} \hat{\bf B})^{ -1}=( \hat{\bf A} - \hat{\bf B} \hat{\bf D}^{ -1} \hat{\bf C})^{ -1} \hat{\bf B} \hat{\bf D}^{ -1},
\end{align}
where, $\hat{\bf A} = {\bf I_{L,L}}$, $\hat{\bf B}={\bf h}^T$, $\hat{\bf C}={\bf h}$ and $\hat{\bf D}=  \dfrac{1}{(a^2 M({\bf S}) + \sigma_{u}^2)} {\boldsymbol{ \Sigma}}_{\epsilon}({\bf S})$, the MMSE Kalman estimation error is rewritten as, 
\begin{align}
\emph{} \nonumber
M({\bf S})  &= \Big(1- \big[1+   {\bf h}^T (a^2 M({\bf S})  + \sigma_{u}^2)   {\boldsymbol{ \Sigma}}_{\epsilon}({\bf S})^{-1}   {\bf h} \big]^{-1} \\ \nonumber
&(a^2 M({\bf S})  + \sigma_{u}^2)  {\boldsymbol{ \Sigma}}_{\epsilon}({\bf S})^{-1}    {\bf h}  \Big)(a^2 M({\bf S})  + \sigma_{u}^2) \\ \label{eq:k_error_2}
&= \dfrac{(a^2 M({\bf S})  + \sigma_{u}^2)}{1+ (a^2 M({\bf S})  + \sigma_{u}^2){\bf h}^T  {\boldsymbol{ \Sigma}}_{\epsilon}({\bf S})^{-1}   {\bf h}}
\end{align}
Letting $\gamma({\bf S}) = {\bf h}^T  {\boldsymbol{ \Sigma}}_{\epsilon}({\bf S})^{-1}   {\bf h}$, \eqref{eq:k_error_2} is reformulated as,
\begin{equation} \label{eq:app_d}
{a^2 M({\bf S}) ^2}{\gamma({\bf S})}+(1+{\sigma_{u}^2}{\gamma({\bf S})}-a^2)M({\bf S}) -\sigma_{u}^2 = 0.
\end{equation}
and therefore,
\begin{align}\label{eq:k_error_3}
\emph{} \nonumber
M({\bf S})  &= \dfrac{\sqrt{\dfrac{(1-a^2)^2}{\gamma({\bf S})^2} + \dfrac{ 2(1-a^2)\sigma_{u}^2+4a^2\sigma_{u}^2}{\gamma({\bf S})}+\sigma_{u}^4}}{(2a^2)} \\
&-{\big( (1-a^2)/\gamma({\bf S})+\sigma_{u}^2   \big)}/{(2a^2)}.
\end{align}
As $\gamma({\bf S})$ increases, we note that the first term of the right hand side is decreasing faster than the increase in the second term in \eqref{eq:k_error_3}. Therefore, the error $M({\bf S}) $ is monotonically decreasing with the increase of  $\gamma({\bf S})$. $M({\bf S}) $ diminishes as $\gamma({\bf S}) \to \infty$. $\gamma({\bf S})$ is always greater than zero by definition. Hence, minimizing $M({\bf S}) $ is equivalent to maximizing $\gamma({\bf S})$.

Since ${\boldsymbol{ \Sigma}}_{\epsilon}({\bf S})$ is diagonal and by substituting $[{\boldsymbol{ \Sigma}}_{\epsilon}({\bf S})]_{l,l}$ as in \eqref{eq:covariance_epsilon}, $\gamma({\bf S})$ is expressed as,
\begin{align}\nonumber
\gamma({\bf S}) &= \sum_{l=1}^L \dfrac{ h_l^2}{[{\boldsymbol{ \Sigma}}_{\epsilon}({\bf S})]_{l,l}} \\ 
\label{eq:k_error_4}
&= \sum_{l=1}^L \dfrac{h_l^2}{\sum_{k=0}^{K} \sum_{b=1}^B \bigg(\sigma_v^2 + \dfrac{\sigma_{x(l)}^2\sigma_{\phi}^2}{g_l \hat{P}_{l,k,b}}\bigg) { s}_{l,k,b} }.
\end{align}
Since $s_{l,k,b} \in \{0,1\}$ and $||{\bf S}_l||_0 = 1$, it is not hard to show that,
\begin{equation}\nonumber
\gamma({\bf S}) = \sum_{l=1}^L \sum_{k=0}^{K} \sum_{b=1}^B \dfrac{h_l^2 {g_l \hat{P}_{l,k,b}} }{ \sigma_v^2 {g_l \hat{P}_{l,k,b}} + {\sigma_{x(l)}^2\sigma_{\phi}^2} } { s}_{l,k,b},
\end{equation}
which proves \eqref{eq:gamma_s}




\bibliographystyle{elsarticle-num}
\bibliography{refs_1,refs_2,refs_3}

\begin{thebibliography}{10}
\expandafter\ifx\csname url\endcsname\relax
  \def\url#1{\texttt{#1}}\fi
\expandafter\ifx\csname urlprefix\endcsname\relax\def\urlprefix{URL }\fi
\expandafter\ifx\csname href\endcsname\relax
  \def\href#1#2{#2} \def\path#1{#1}\fi

\bibitem{Bushnaq2017EH}
O.~M. Bushnaq, T.~Y. Al-Naffouri, S.~P. Chepuri, G.~Leus, Joint sensor
  placement and power rating selection in energy harvesting wireless sensor
  networks, in: 2017 25th European Signal Processing Conference (EUSIPCO),
  2017, pp. 2423--2427.
\newblock \href {http://dx.doi.org/10.23919/EUSIPCO.2017.8081645}
  {\path{doi:10.23919/EUSIPCO.2017.8081645}}.

\bibitem{Bushnaq2017dynamic}
O.~M. Bushnaq, A.~Chaaban, T.~Al-Naffouri, Joint sensor location/power rating
  optimization for temporally-correlated source estimation, in: 2017 IEEE 18th
  International Workshop on Signal Processing Advances in Wireless
  Communications (SPAWC), 2017, pp. 1--5.
\newblock \href {http://dx.doi.org/10.1109/SPAWC.2017.8227640}
  {\path{doi:10.1109/SPAWC.2017.8227640}}.

\bibitem{Zanella2014IoT}
A.~Zanella, N.~Bui, A.~Castellani, L.~Vangelista, M.~Zorzi, Internet of things
  for smart cities, IEEE Internet of Things Journal 1~(1) (2014) 22--32.
\newblock \href {http://dx.doi.org/10.1109/JIOT.2014.2306328}
  {\path{doi:10.1109/JIOT.2014.2306328}}.

\bibitem{Xu2014IoT}
L.~D. Xu, W.~He, S.~Li, Internet of things in industries: A survey, IEEE
  Transactions on Industrial Informatics 10~(4) (2014) 2233--2243.
\newblock \href {http://dx.doi.org/10.1109/TII.2014.2300753}
  {\path{doi:10.1109/TII.2014.2300753}}.

\bibitem{Lin2017IoT}
C.~Lin, D.~Deng, a.~L.~Lu, Many-objective sensor selection in {IoT} systems,
  IEEE Wireless Communications 24~(3) (2017) 40--47.
\newblock \href {http://dx.doi.org/10.1109/MWC.2017.1600409}
  {\path{doi:10.1109/MWC.2017.1600409}}.

\bibitem{greedy2010}
M.~Shamaiah, S.~Banerjee, H.~Vikalo, Greedy sensor selection: Leveraging
  submodularity, in: Decision and Control (CDC), 2010 49th IEEE Conference on,
  IEEE, 2010, pp. 2572--2577.

\bibitem{greedy2015sundeep}
S.~Rao, S.~P. Chepuri, G.~Leus, Greedy sensor selection for non-linear models,
  in: Computational Advances in Multi-Sensor Adaptive Processing (CAMSAP), 2015
  IEEE 6th International Workshop on, IEEE, 2015, pp. 241--244.

\bibitem{boyd2009}
S.~Joshi, S.~Boyd, Sensor selection via convex optimization, IEEE Transactions
  on Signal Processing 57~(2) (2009) 451--462.

\bibitem{boyd2004convex}
S.~Boyd, L.~Vandenberghe, Convex optimization (2004).

\bibitem{chepuri2015sparsity}
S.~P. Chepuri, G.~Leus, Sparsity-promoting sensor selection for non-linear
  measurement models, IEEE Transactions on Signal Processing 63~(3) (2015)
  684--698.

\bibitem{sundeep16FnT}
S.~Chepuri, G.~Leus, Sparse sensing for statistical inference, Foundations and
  Trends in Signal Processing 9~(3) (2016) 233--368.

\bibitem{Giannakis2012sensor_censoring}
E.~J. Msechu, G.~B. Giannakis, Sensor-centric data reduction for estimation
  with {WSN}s via censoring and quantization, IEEE Transactions on Signal
  Processing 60~(1) (2012) 400--414.

\bibitem{jawaid2015submodularity}
S.~T. Jawaid, S.~L. Smith, Submodularity and greedy algorithms in sensor
  scheduling for linear dynamical systems, Automatica 61 (2015) 282--288.

\bibitem{savage2009optimal}
C.~O. Savage, B.~F. La~Scala, Optimal scheduling of scalar {G}auss-{M}arkov
  systems with a terminal cost function, IEEE Transactions on Automatic Control
  54~(5) (2009) 1100--1105.

\bibitem{wang2013sequential}
G.~Wang, J.~Chen, J.~Sun, On sequential {K}alman filtering with scheduled
  measurements, in: Cyber Technology in Automation, Control and Intelligent
  Systems (CYBER), 2013 IEEE 3rd Annual International Conference on, IEEE,
  2013, pp. 450--455.

\bibitem{sparsity2016spain}
M.~Calvo-Fullana, J.~Matamoros, C.~Ant{\'o}n-Haro, S.~M. Fosson,
  Sparsity-promoting sensor selection with energy harvesting constraints, in:
  Acoustics, Speech and Signal Processing (ICASSP), 2016 IEEE International
  Conference on, IEEE, 2016, pp. 3766--3770.

\bibitem{zhang2015sensor}
H.~Zhang, R.~Ayoub, S.~Sundaram, Sensor selection for optimal filtering of
  linear dynamical systems: Complexity and approximation, in: IEEE Conference
  on Decision and Control (CDC), 2015.

\bibitem{zhang2017sensor}
H.~Zhang, R.~Ayoub, S.~Sundaram, Sensor selection for {K}alman filtering of
  linear dynamical systems: Complexity, limitations and greedy algorithms,
  Automatica.

\bibitem{ren2014dynamic}
Z.~Ren, P.~Cheng, J.~Chen, L.~Shi, H.~Zhang, Dynamic sensor transmission power
  scheduling for remote state estimation, Automatica 50~(4) (2014) 1235--1242.

\bibitem{citeulike:347166}
R.~E. Kalman,
  \href{http://www.cs.unc.edu/\~{}welch/kalman/media/pdf/Kalman1960.pdf}{{A New
  Approach to Linear Filtering and Prediction Problems}}, Transactions of the
  ASME – Journal of Basic Engineering~(82 (Series D)) (1960) 35--45.
\newline\urlprefix\url{http://www.cs.unc.edu/\~{}welch/kalman/media/pdf/Kalman1960.pdf}

\bibitem{summers2016actuator}
T.~Summers, Actuator placement in networks using optimal control performance
  metrics, in: Decision and Control (CDC), 2016 IEEE 55th Conference on, IEEE,
  2016, pp. 2703--2708.

\bibitem{calvo2016sensor}
M.~Calvo-Fullana, J.~Matamoros, C.~Ant{\'o}n-Haro, Sensor selection and power
  allocation strategies for energy harvesting wireless sensor networks, IEEE
  Journal on Selected Areas in Communications 34~(12) (2016) 3685--3695.

\bibitem{Shirazi2017Digital}
M.~Shirazi, A.~Sani, A.~Vosoughi, Sensor selection and power allocation via
  maximizing {B}ayesian {F}isher information for distributed vector estimation,
  in: 2017 51st Asilomar Conference on Signals, Systems, and Computers, 2017,
  pp. 1379--1383.
\newblock \href {http://dx.doi.org/10.1109/ACSSC.2017.8335580}
  {\path{doi:10.1109/ACSSC.2017.8335580}}.

\bibitem{Shirazi2017fisher}
M.~Shirazi, A.~Vosoughi, {F}isher information maximization for distributed
  vector estimation in wireless sensor networks (2017).
\newblock \href {http://arxiv.org/abs/1705.00803} {\path{arXiv:1705.00803}}.

\bibitem{kay1993fundamentals}
S.~M. Kay, Fundamentals of statistical signal processing: estimation theory.

\bibitem{cover2006elements}
T.~Cover, J.~Thomas, Elements of Information Theory.

\bibitem{anderson1979optimal}
B.~D. Anderson, J.~B. Moore, Optimal filtering, Englewood Cliffs 21 (1979)
  22--95.

\bibitem{cvx}
M.~Grant, S.~Boyd, {CVX}: Matlab software for disciplined convex programming,
  version 2.1, http://cvxr.com/cvx (Mar. 2014).

\end{thebibliography}


\end{document}